\documentclass[numberedappendix]{emulateapj}

\newcommand{\swift}{{\it Swift}}
\newcommand{\chandra}{{\it Chandra}}
\newcommand{\xmm}{{\it XMM-Newton}}
\newcommand{\TBDtot}{12} 
\newcommand{\TBDrec}{9} 
\newcommand{\TBDold}{3} 
\newcommand{\TBDfull}{5} 
\newcommand{\TBDmarg}{3} 
\newcommand{\TBDnon}{4} 
\newcommand{\TBDSSS}{2} 
\newcommand{\TBDobsnum}{27} 

\begin{document}

\title{Swift X-ray Observations of Classical Novae}

\author{J.-U. Ness\altaffilmark{1}, G.J. Schwarz\altaffilmark{2},
A. Retter\altaffilmark{3}, S. Starrfield\altaffilmark{1},
 J.H.M.M. Schmitt\altaffilmark{4}, N. Gehrels\altaffilmark{5},
D. Burrows\altaffilmark{6}, J.P. Osborne\altaffilmark{7}}

\altaffiltext{1}{School of Earth and Space Exploration, Arizona
State University, Tempe, AZ 85287-1404, USA: Jan-Uwe.Ness,sumner.starrfield@asu.edu}
\altaffiltext{2}{Department of Geology \& Astronomy, West Chester University, 750 S. Church Street, West Chester, PA 19383, USA}
\altaffiltext{3}{Department of Astronomy and Astrophysics, Penn State University, 525 Davey Lab, University Park, PA, 16802-6305, USA}
\altaffiltext{4}{Hamburger Sternwarte, Gojenbergsweg 112, 21029
Hamburg, Germany}
\altaffiltext{5}{NASA/Goddard Space Flight Center, Greenbelt, Maryland 20771, USA}
\altaffiltext{6}{Department of Astronomy and Astrophysics, Penn State University, University Park, Pennsylvania 16802, USA}
\altaffiltext{7}{Department of Physics \& Astronomy, University of Leicester, Leicester, LE1 7RH, UK}

\begin{abstract}

The new $\gamma$-ray burst (GRB) mission \swift\ has obtained pointed
observations of several classical novae in outburst. We analyzed all the
observations of classical novae from the \swift\ archive up to 30
June, 2006. We analyzed usable observations of \TBDtot\
classical novae and found \TBDnon\ non-detections, \TBDmarg\ weak
sources and \TBDfull\ strong sources. This includes detections of \TBDSSS\
novae exhibiting spectra resembling those of Super Soft X-ray binary Source spectra
(SSS) implying ongoing nuclear burning on the white dwarf surface. With these new
\swift\ data, we add to the growing statistics of the X-ray duration and
characteristics of classical novae.

\end{abstract}


\date{accepted by ApJ (March 8, 2007)\footnote{This
version contains additional material}}

\section{Introduction}
\label{intro}

The explosions of Classical Novae (CNe) occur in close binary star systems where
mass is accreted onto a white dwarf (WD) from a low-mass secondary star. The accreted
material gradually becomes degenerate, and when temperatures become high enough and
the pressure at the bottom of the accreted envelope exceeds a certain value,
a thermonuclear runaway results. Enough energy is deposited in the accreted
material to eject some fraction from the WD. The outburst can last
several months to years,
and CNe become active in X-rays at some time during their outburst
\citep[e.g., ][]{pietsch05,pietsch06}. The relatively
slow evolution makes CNe ideal targets for \swift, as no accurate times of
observation need to be scheduled and they can be observed with great flexibility.

The evolution of X-ray emission starts soon after outburst when the expanding nova
envelope is still small and dense. The energy peak during this earliest time,
called the ``fireball'' phase \citep{shore94,schwarz01}, is
expected to occur at X-ray
wavelengths. This phase is extremely difficult to observe since
it is predicted to last only a few hours after the beginning of the outburst
so that it is over before the nova can be discovered in the optical. As the
envelope increases in size and cools, the opacity increases, shifting the
radiation towards wavelengths longer than the hydrogen absorption edge at
13.6\,eV (911\,\AA). The expanding, cooling envelope gives rise to the large
and rapid increase in the visual luminosity of the nova. Around the time of
visual maximum, X-rays from the underlying WD are trapped inside the large
column density of the ejecta. However, shocks may develop within the early
expanding ejecta producing a hard, sub-Eddington luminosity, X-ray spectral energy
distribution \citep[{\it e.g.}, V382\,Vel,][]{Orio2001,mukai01}. As the shell
continues to expand, the density and opacity drop, and X-rays emitted from the surface
of the WD eventually become visible \citep{kr96}. Soft X-ray emission
originates from hot layers produced by the nuclear burning of the remaining accreted
material on the WD. This material burns in equilibrium at a high but constant
luminosity, $\propto$ L$_{\rm Eddington}$ \citep{gallagher78}. The
spectral energy distribution resembles that of the
super-soft X-ray sources (SSS), \citep[{\it e.g.} {Cal 83},][]{paerels01,lanz04}.

Recent observations of novae during this phase show that the flux can also be
highly variable on short timescales \citep[see, {\it e.g.}][]{v4743,osborne06}.
The duration of this SSS phase is predicted to be inversely proportional to the
WD mass \citep{starrfield91}, but a review of the ROSAT X-ray sky survey
\citep{orio01} showed that the timescale is much shorter than predicted,
implying either that the masses of WDs in CNe are much higher than
commonly assumed, or that the X-ray turn-off is a function of more than the WD mass.
A different approach was proposed by \cite{greiner03} who found from
the available data that systems with
shorter orbital periods display long durations of supersoft X-ray phases, while
long-period systems show very short or no SSS phase at all. They speculate that
shorter periods may be related to a higher mass transfer rate (e.g., by increased
irradiation) and thus to the amount of material accreted before the explosion.

Another source of X-ray emission from novae are emission lines from material that
has been radiatively ionized. These lines are expected to be present during the
entire evolution, but are usually only observed once the nuclear fuel is consumed
and the bright continuous SSS spectrum (that outshines the emission lines) fades.
The ejecta
quickly recombine until a collisional equilibrium
is reached. This equilibrium reflects the kinetic
temperature distribution of the plasma and the elemental abundances can be derived
\citep{ness_vel}. This phase is called the nebular phase, and is also seen at other
wavelengths \citep{shore03}. In some cases, X-ray emission from lines has been
observed earlier, either prior to the SSS phase, superimposed on the SSS spectrum
({\it e.g.}, V1494\,Aql), or during
times of extreme variability when the SSS emission suddenly declines before
recovery \citep[{\it e.g.}, V4743\,Sgr,][]{v4743}.\\

X-ray observations provide important clues to the properties and
dynamics of novae. The evolution of the SSS depends on the WD mass, the mass
loss rate (via radiatively driven winds or common envelope mass loss), the
amount of ejected material, the amount of material remaining on the WD, the
binary separation, and the expansion velocity of the ejecta. Analysis of the X-ray
emission can also provide insight into the composition of the ejecta.
Unfortunately, only a few novae have been studied extensively in X-rays with
much of the recent information coming from \chandra\ and \xmm.
Now, with \swift, we have the possibility of studying novae in
large numbers in order to assess statistical properties. \swift\ also
allows for coordination of higher-resolution observations with \chandra\
and \xmm\ since it has always been difficult to predict the brightness in X-rays
and to estimate the optimal exposure time for X-ray observations with the gratings.

In this paper we discuss the \swift\ XRT instrument briefly
in Sect.~\ref{swift}. Since interstellar absorption
plays an important role in Galactic novae, we discuss the expected
effects on the detection of SSS in Sect.~\ref{nh}.
In Sect.~\ref{analysis} we sort the observations
by non-detections, detections of weak sources, and strong sources
based on our extraction statistics detailed out in the appendix section
\ref{extraction}. We discuss our results in 
Sect.~\ref{discuss} and expand on the type of X-ray emission that
was detected in the \swift\ observations.
Finally, Appendix \ref{atargets}
provides background information for each nova.

\section{\swift\ Observations}
\label{obs}

\subsection{The instrument}
\label{swift}

\begin{figure}
\resizebox{\hsize}{!}{\includegraphics{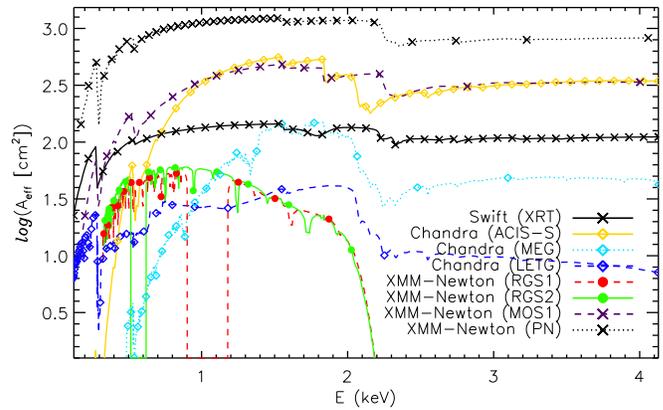}}
\caption{\label{inscmp}Comparison of XRT effective areas of \swift\ with other
X-ray instruments that are sensitive in the same spectral region. None of the \chandra\
or \xmm\ gratings has a higher effective area than the XRT. The CCD detectors:
ACIS-S (\chandra) and EPIC (PN/MOS1; \xmm) have higher effective areas. The
spectral resolution of the XRT is similar to either the ACIS or
EPIC detectors, but the spectral resolution of the gratings is much superior.}
\end{figure}

Our entire dataset has been obtained with the X-ray Telescope (XRT) aboard \swift\
\citep{B05}. The XRT instrument is a
CCD detector behind a Wolter Type I grazing incidence mirror consisting of 12 nested
shells. The field of view covers $23.6\arcmin \times 23.6$\arcmin, imaged on a
detector with $600\times 600$ pixels, thus each pixel corresponds to 2.36\arcsec.
 The point spread function (PSF) can be parameterized, and
source radii of 10\,pixels and 5\,pixels include 80.5\,percent and 60.5\,percent
of the total energy, respectively \citep{moretti04}. The positional
accuracy is 2.5\arcsec, or one pixel. The
energy range covers $0.2-10$\,keV with the Full Width at Half Maximum (FWHM)
energy resolution
varying from $\sim50$\,eV at 0.1\,keV to $\sim190$\,eV at 10\,keV.
In Fig.~\ref{inscmp} we show a comparison of effective areas of \swift,
\chandra, and \xmm. In general, the \swift\ XRT has a smaller effective area than the
\chandra /ACIS-S except at low energies (long wavelengths) where the \chandra\ ACIS-S
suffers from large calibration uncertainties. The EPIC detectors PN and MOS1
(\xmm) have 50\,percent and 20\,percent higher effective areas, respectively.
The XRT, ACIS, and EPIC instruments have similar spectral resolution. The XRT
effective area is larger than that of the gratings aboard \chandra\ (LETG and
MEG) and \xmm\ (RGS1 and RGS2), but the XRT spectral resolution is much lower.
Fig.~\ref{inscmp} demonstrates that the XRT is an ideal instrument for exploring
the emission level of novae using reasonable exposure times ($\sim3-6$\,ksec).
Once a nova is detected with \swift\ and found sufficiently
bright, additional observations can be requested with
the high-resolution grating instruments aboard \chandra\ and \xmm\ to obtain detailed
spectral information.
The detector was operated in Photon Counting (PC) mode which
provides 2D imaging, spectral information, and 2.5-sec time resolution.

\subsection{Observability of the Super Soft Source phase in Classical Novae}
\label{nh}

\begin{figure}
\resizebox{\hsize}{!}{\includegraphics{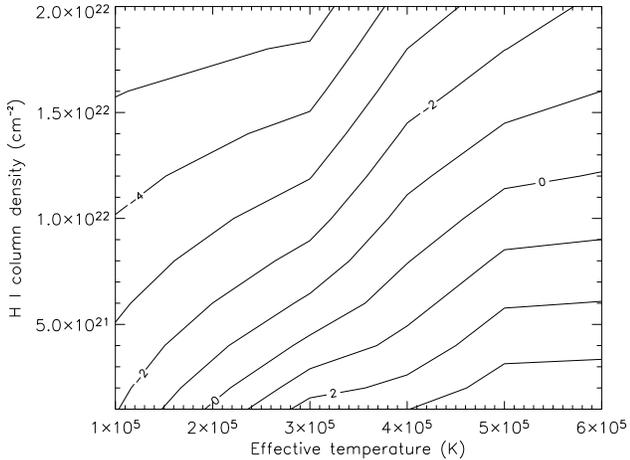}}
\caption{\label{tvsn}Contour plot of the log$_{10}$ Swift counts predicted
with a 1-ksec observation from a SSS a a function of temperature
(Cloudy model) and interstellar extinction (N$_{\rm H}$) for a
source at a distance of 1\,kpc.}
\end{figure}

Unfortunately, many Galactic CNe have large extinction rates which has a
pronounced effect on the observability of the SSS emission. To illustrate this
we attenuate a series of {\it Cloudy} models \citep{Fer98} through different
amounts of \ion{H}{1} column densities and determine the resulting \swift\
counts. The models use non-LTE atmospheres by \cite{rauch97} for planetary
nebula nuclei with effective temperatures in the range of
$(1-6)\times 10^5$\,K (8--50eV) as the input source. The model luminosity was fixed
at $10^{38}$\,erg\,sec$^{-1}$, which is approximately the Eddington luminosity
for a 1-M$_\odot$ star. The spectral energy distributions were
attenuated by \ion{H}{1} column densities ranging from
$(1-20)\times 10^{21}$\,cm$^{-2}$ and then converted to \swift\ net count rates
using the effective area of the XRT and scaled to a 1\,ksec exposure at
1\,kpc. The logarithms of the expected total XRT counts are shown as contours
in Fig.~\ref{tvsn}. As expected, the ability to detect a SSS is highly dependent
on N$_{\rm H}$ and the source's effective temperature. Sources with low temperatures
are more difficult to detect if significant N$_{\rm H}$-absorption takes place.
Without long ($>10$\,ksec) exposures, N$_{\rm H}\sim 10^{22}$\,cm$^{-2}$ is the
highest number that allows a detection of SSS emission at 1\,kpc distance
with \swift. Since most novae are further away than 1\,kpc, the situation is
even worse.

\subsection{Targets}
\label{targets}

Although primarily a mission to study the temporal evolution of $\gamma$-ray
bursts (GRB), \swift\ also provides Targets of Opportunity (ToO) observations
for non-GRB events. Through ToO observations we have obtained X-ray 
observations of \TBDrec\ recent ($< 1-2$ years since outburst) and \TBDold\ 
older ($> 3$ years) CNe. We consider all observations, although those of 
\object[V1188 Sco]{V1188\,Sco} on 
\dataset[ADS/Sa.SWIFT#o/00035191001]{21 May, 2006} and of V574\,Pup taken on
\dataset[ADS/Sa.SWIFT#o/00035170004]{26 July, 2005} were extremely short 
with hardly any signal.
Table \ref{novaparameters} provides a list of important
physical parameters associated with these novae including the maximum visual 
magnitudes, the date of discovery, the time to decline 2 magnitudes from 
maximum (t$_2$), the early measured expansion velocities, E($B-V$) reddening 
values, H\,{\sc i} column densities, and distances.
In the appendix \ref{atargets} we summarize background information 
for each nova in our sample.

\begin{deluxetable*}{lrcrrrcrr}
\tablecaption{\label{novaparameters}Basic nova parameters}
\tabletypesize{\footnotesize} 
\tablehead{\colhead{Object} & \colhead{V$_{max}$} & \colhead{Date\tablenotemark{a}} &
\colhead{t$_2$\tablenotemark{b}} &
\colhead{v$_{\rm exp}$\tablenotemark{c}} & \colhead{E(B--V)} & 
\colhead{N$_H$\tablenotemark{d}} & \colhead{Dis.} &
\colhead{Refs.\tablenotemark{e}} \\
\colhead{} & \colhead{(mag)} & \colhead{} & \colhead{(days)} &
\colhead{(km\,sec$^{-1}$)} & \colhead{(mag)} &
\colhead{(cm$^{-2}$)} & \colhead{(kpc)}
}
\startdata
V574\,Pup & 8    & 11/26/2004 &          13 & 650H,860P   & 0.5           & 6.5e21 &  3.20 & 1,2,3 \\
V382\,Nor & 9    & 03/19/2005 &          13 & 1100P       & 0.6-1.1       & 1.8e22 & 13.80 & 3,3,3 \\
V1663\,Aql & 10.5 & 06/10/2005 &          16 & 700P        & $\sim$2       & 1.8e22 & 5.5$\pm$1 & 4,5,4 \\
V5116\,Sgr & $<$8 & 07/04/2005 &          20 & 1300P       & 0.24$\pm$0.08 & 1.6e21 & 11.30& 3,3,3 \\
V1188\,Sco & 8.9  & 07/26/2005 &          12 & 1730H,4000Z & $\sim$1?      & 5.0e21 & 7.5 & 3,3,3 \\
V1047\,Cen & 8.83 & 09/04/2005 &           ? & 850H        & 3.3?          & 1.6e22 & ? & -,3,- \\
V476\,Sct & 10.9 & 09/30/2005 &          15 & 1200H       & 1.9$\pm$0.1   & 1.1e22 & 4$\pm$1 & 6,6,6 \\
V477\,Sct & 10.4 & 11/10/2005 &           3 & 2700H,6000Z & $>$1.3        & 4.8e21 & 11 & 7,7,7 \\
LMC\,2005 & 12.6 & 11/27/2005 &           ? & --          & 0.15          & 6.3e20 & 50 & 3,3,3 \\
V723\,Cas & 7.1  & 12/17/1995 &        slow & ~700H       & $\sim$0.5     & 2.4e21 & 4 & -,8,9 \\
V1494\,Aql & 4    & 12/03/1999 & 6.6$\pm$0.5 & 1300H,1850P & 0.6$\pm$0.1   & 4.2e21 & 1.6$\pm$0.2& 10,-,11 \\
V4743\,Sgr & 5    & 09/20/2002 &           9 & 2400H       & low?          & 1.4e21 & 6.3 & 3,-,12 \\
\enddata
\tablenotetext{a}{Date of visual maximum (mm/dd/yyyy).}
\tablenotetext{b}{Time to decline 2 magnitudes from maximum.}
\tablenotetext{c}{Expansion velocity. Taken from early IAU circulars. Trailing letters
indicate how the velocity was measured: H = Full-Width at Half-Maximum, P = P Cygni
absorption, Z = Full-Width at Zero-Intensity.}
\tablenotetext{d}{The H\,{\sc i} column densities were obtained from the Heasarc
NH tool. They are the line-of-sight column densities through the entire Galaxy
toward the coordinates of each object.}
\tablenotetext{e}{References for t$_2$, E(B--V), and Distance:\\
1 = \cite{2005IBVS.5638....1S};
2 = R.J. Rudy 2006, private communication;
3 = t$_2$: from AAVSO light curve and IAU circulars. E(B--V): using \citet{vandenbergh87}
    and the (B--V) color evolution or assuming E(B--V) $\sim$ N$_H$/4.8$\times$10$^{21}$ cm$^{-2}$.
    Distance: using the M$_V$ vs $t_2$ relationship of \citet{DVL95};
4 = \citet{lane06};
5 = \citet{2005IAUC.8640....2P};
6 = \citet{Munari06b},\citet{2005IAUC.8638....1P};
7 = \citet{Munari06a};
8 = \citet{1996A&A...315..166M};
9 = \citep{Iijima98};
10 = \citet{kissth00};
11 = \citet{Iijima03};
12 = \citet{lyke}.
}
\end{deluxetable*}

\section{Analysis of the \swift\ XRT data}
\label{analysis}

Currently the \swift\ archive contains \TBDobsnum\ observations of \TBDtot\
novae obtained with the X-ray Telescope (XRT) instrument
(see Sect.~\ref{swift}). We analyzed the data using the \swift\ reduction
packages in HEAsoft version
6.06\footnote{http://swift.gsfc.nasa.gov/docs/software/lheasoft/} and the
latest calibration data (version 20060427). We started with the photon event
files (level 2) that are available in the public \swift\ archive and carried
out three statistical tests to determine if the sources were detected. We
then calculated count rates or upper limits in cases where no detections can
be claimed. The first criterion is the formal detection likelihood from
comparison of the number of counts measured in the source extraction region
(circular with 10\,pixels radius around expected sky position) with the
background extraction region (annulus around source extraction region with
inner radius 10\,pixels and outer radius of 100\,pixels). We next
determined if there was
a concentration of counts near the center of the source extraction
region. Finally, we compared the spectral distributions of
photons in the source and background extraction regions, but
this is only a soft criterion, as similar spectral distributions
do not rule out the presence of a source. We present our
statistical methods in more detail in the appendix
Sect.~\ref{extraction} to justify our tests.

In Table~\ref{obstab} we provide all the
\swift\ observational data including observation date,
time since visual maximum, exposure time, number of counts in the source
and background extraction regions, and either the net source count rate per detect cell
with $1\sigma$ uncertainty, or the upper limit.

\begin{deluxetable}{crrrrcl}
\tabletypesize{\scriptsize}
\tablewidth{0pt}
\tablecolumns{7}
\tablecaption{\label{obstab}Measured count rates in our sample}
\tablehead{
\colhead{Start} & \colhead{$\Delta$t\tablenotemark{a}} &
\colhead{Exp.} & \colhead{Src.} & \colhead{Bg.} &
\colhead{Rate\tablenotemark{b,d}} & \colhead{$R_{1/2}$\tablenotemark{c}} \\
\colhead{Date} & \colhead{(d)} & \colhead{(ks)} & \colhead{(cts)\tablenotemark{b}} &
\colhead{(cts)\tablenotemark{b}} & \colhead{(cts/ks)} & \colhead{(\%)}}
\startdata
\cutinhead{\bf V574\,Pup}
\dataset[ADS/Sa.SWIFT#o/00035170001]{05/20/2005} & 175 & 1.0 & 10 & 0.4  & $12.0\,\pm\,4.0$ & 60\\
\dataset[ADS/Sa.SWIFT#o/00035170003]{05/25/2005} & 180 & 1.9 & 22 & 0.4  & $14.0\,\pm\,3.0$ & 77\\
\dataset[ADS/Sa.SWIFT#o/00035170004]{05/26/2005} &     & 0.01  & 0  & 0    & 0 & 0\\
\dataset[ADS/Sa.SWIFT#o/00035170005]{07/29/2005} & 245 & 1.1 & 15 & 0.3  & $16.0\,\pm\,4.0$ & 67\\
\dataset[ADS/Sa.SWIFT#o/00035170006]{07/30/2005} & 246 & 7.1 & 51 & 2.1  & $8.5\,\pm\,1.3$  & 73\\
\dataset[ADS/Sa.SWIFT#o/00035170007]{08/06/2005} & 253 & 7.8 & 57 & 2.0  & $8.7\,\pm\,1.2$  & 67\\
\dataset[ADS/Sa.SWIFT#o/00035170008]{08/09/2005} & 256 & 2.2 & 22 & 0.6  & $12.0\,\pm\,3.0$ & 64\\
\dataset[ADS/Sa.SWIFT#o/00035170009]{08/10/2005} & 257 & 2.8 & 18 & 0.8  & $7.8\,\pm\,2.0$  & 67\\
\dataset[ADS/Sa.SWIFT#o/00035170010]{08/11/2005} & 258 & 1.8 & 11 & 0.5  & $7.3\,\pm\,2.5$  & 56\\
\dataset[ADS/Sa.SWIFT#o/00035170011]{08/17/2005} & 264 & 4.7 & 34 & 1.2  & $8.6\,\pm\,1.6$  & 79\\
\cutinhead{\bf V382\,Nor}
\dataset[ADS/Sa.SWIFT#o/00035195002]{01/26/2006} & 313 & 6.1 & 72 & 4.0  & $14.0\,\pm\,2.0$ & 75\\
\cutinhead{\bf V1663\,Aql}
\dataset[ADS/Sa.SWIFT#o/00035193001]{08/15/2005} &  66 & 1.3 & 9  & 0.3  & $8.6\,\pm\,3.3$  & 78\\
\dataset[ADS/Sa.SWIFT#o/00035193002]{03/04/2006} & 267 & 6.5 & 3  & 1.1  & $<0.5\tablenotemark{d}$  & 0 \\
\cutinhead{\bf V5116\,Sgr}
\dataset[ADS/Sa.SWIFT#o/00035192001]{08/29/2005} &  56 & 3.1 &  5 & 2.0  & $1.2\,\pm\,1.0$  & 80\\
\cutinhead{\bf V1188\,Sco}
\dataset[ADS/Sa.SWIFT#o/00035191001]{05/21/2006} & 98 & 0.2  & 1  & 0.1  & $<7\tablenotemark{d}$  & 0\\
\dataset[ADS/Sa.SWIFT#o/00035191002]{06/17/2006} & 124 & 5.0 & 3  & 2.3  & $<9.1\tablenotemark{d}$  & 33\\
\cutinhead{\bf V1047\,Cen}
\dataset[ADS/Sa.SWIFT#o/00035231001]{09/11/2005} &  66 & 3.9 & 19 & 1.3  & $5.6\,\pm\,1.5$  & 53\\
\dataset[ADS/Sa.SWIFT#o/00035231002]{01/23/2006} & 141 & 5.3 & 53 & 3.1  & $11.0\,\pm\,2.0$ & 72\\
\cutinhead{\bf V476\,Sct}
\dataset[ADS/Sa.SWIFT#o/00035229001]{02/12/2006} & 135 & 3.8 & 2  & 1.0  & $<8.2\tablenotemark{d}$  & 50\\
\cutinhead{\bf V477\,Sct}
\dataset[ADS/Sa.SWIFT#o/00035230001]{03/07/2006} & 117 & 6.2 & 15 & 1.8  & $2.7\,\pm\,0.8$  & 87\\
\dataset[ADS/Sa.SWIFT#o/00035230002]{03/15/2006} & 125 & 4.7 &  8 & 1.3  & $1.8\,\pm\,0.8$  & 50\\
\cutinhead{\bf LMC\,2005}
\dataset[ADS/Sa.SWIFT#o/00030348001]{12/02/2005} & 5   & 0.1 & 0  & 0.1  & $<1.0\tablenotemark{d}$ & 0 \\
\dataset[ADS/Sa.SWIFT#o/00030348002]{12/01/2005} & 4   & 4.1 & 1  & 1.3  & $<8.6\tablenotemark{d}$ & 100\\
\dataset[ADS/Sa.SWIFT#o/00030348003]{12/02/2005} & 5   & 3.9 & 2  & 1.2  & $<8.7\tablenotemark{d}$ & 100\\
\dataset[ADS/Sa.SWIFT#o/00030348004]{12/03/2005} & 6   & 5.7 & 1  & 1.7  & $<7.0\tablenotemark{d}$ & 50 \\
\cutinhead{\bf V723\,Cas}
\dataset[ADS/Sa.SWIFT#o/00030361001]{01/31/2006} & 3698 & 6.9 & 147 & 1.7 & $26.0\,\pm\,2.0$ & 69\\
\cutinhead{\bf V1494\,Aql}
\dataset[ADS/Sa.SWIFT#o/00035222001]{03/10/2006} & 2289 & 1.4 & 0 &  0.3 & $<1.4\tablenotemark{d}$ & 0\\
\dataset[ADS/Sa.SWIFT#o/00035222002]{03/13/2006} & 2292 & 1.2 & 0 &  0.2 & $<1.4\tablenotemark{d}$ & 0\\
\dataset[ADS/Sa.SWIFT#o/00035222003]{05/19/2006} & 2359 & 2.5 & 1 &  0.5 & $<1.0\tablenotemark{d}$ & 100\\
\cutinhead{\bf V4743\,Sgr}
\dataset[ADS/Sa.SWIFT#o/00035221001]{03/08/2006} & 1265 & 5.8 & 267 & 2.3 & $27.0\,\pm\,2.0$ & 70\\
\enddata
\tablenotetext{a}{Interval between visual maximum and the XRT observation in days}
\tablenotetext{b}{per detect cell (10\,pixels radius); count rates with $1\sigma$ uncertainties}
\tablenotetext{c}{fraction of counts within 5\,pixels (\%)}
\tablenotetext{d}{95.5-\% upper limit}
\end{deluxetable}

\subsection{Non-detections, failure of all criteria}

Only upper limits could be established for the observations
of V476\,Sct, V1188\,Sco, LMC\,2005, and V1494\,Aql.
\begin{figure*}
\resizebox{\hsize}{!}{\includegraphics{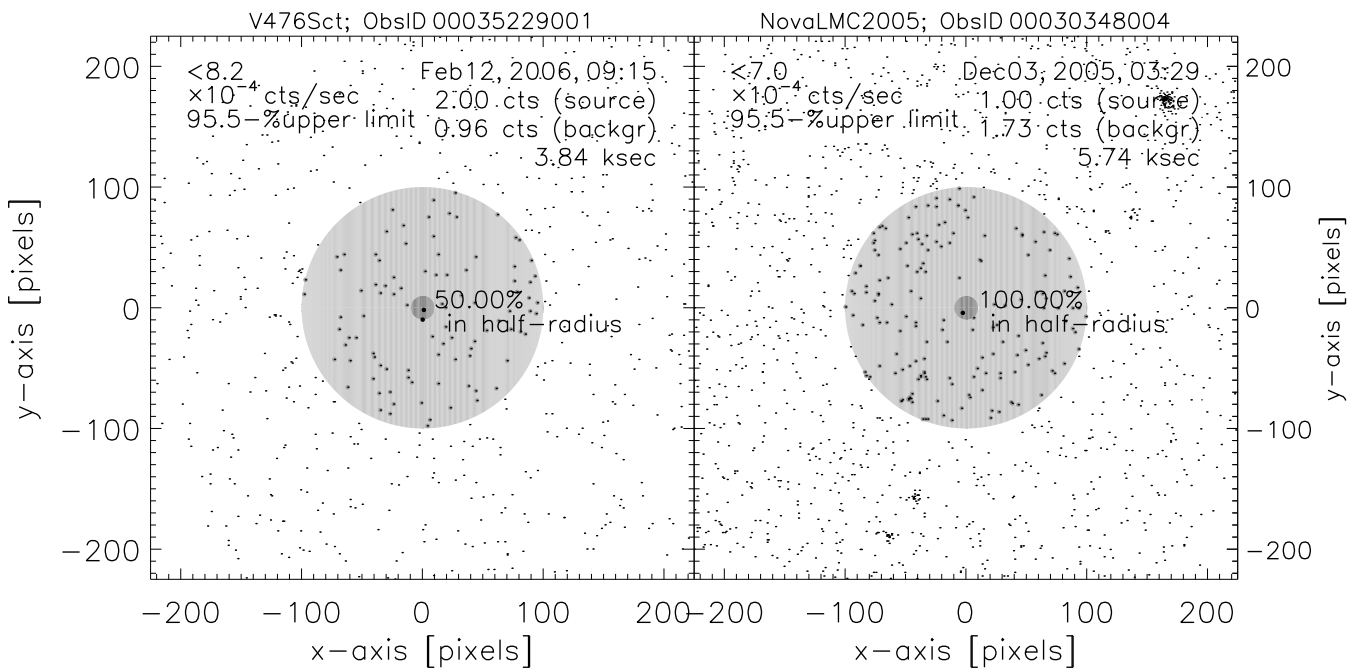}}
\caption{\label{firstphot}V476\,Sct (left) and LMC\,2005 (right) were not
detected. We show the recorded photon positions and mark the source
extraction region with dark shading and the background extraction region
(annulus with inner radius 10\,pixels and outer radius 100\,pixels) with
light shading. The pixel- and sky coordinates are given in the bottom.
The upper limits are given in the upper left. Observation date,
extraction radius, number of source- and background counts per detect cell,
and exposure time are given in the upper right. The percentage of source counts
that reside within 25\,percent area (radius 5\,pixels) is given to the right
of each source extraction region to give an impression of the concentration
towards the center within the source extraction region.}
\end{figure*}

 In Fig.~\ref{firstphot} we show the photon plot
centered around the expected source position of V476\,Sct and LMC\,2005. The
extraction regions for source and background,
as defined in Sects.~\ref{analysis} and \ref{extraction} are marked with dark
and light gray colors, respectively, and the counts within the source
extraction region are plotted with small black bullets. In the upper
right corner we list: the date of observation, source extraction region,
the number of counts in the respective extraction regions, and the
exposure times. In the upper left corner we display the source net count rate
(or the 95.5-\% upper limit if no significant excess count rate is found) as
well as the detection likelihood, the number of net source- and background counts
$S$ and $B$, respectively, as defined in Eq.~\ref{model}.
 For V1188\,Sco, all criteria fail
for all observations, and the source is clearly not detected.
A similar case is encountered for all three observations of LMC\,2005,
where we show the last observation in the right panel of Fig.~\ref{firstphot}.
Further, we found no significant detections in all three observations of
V1494\,Aql.

We found only a marginal detection in the \chandra\
observation of V1494\,Aql five years prior to the \swift\ observations
(see Sect.~\ref{supext}).
We used the proposal planning tool PIMMS in order to convert the \chandra\
count rate into an expected \swift\ XRT count rate. With various
model assumptions PIMMS predicts more than
0.04 XRT counts per second, while we measured less than 0.0014
counts per second
in each observation (Table~\ref{obstab}). This implies that the
source has faded by at least a factor 30 within five years. 

\subsection{Weak sources $<0.01$\,cps}

\begin{figure*}
\resizebox{\hsize}{!}{\includegraphics{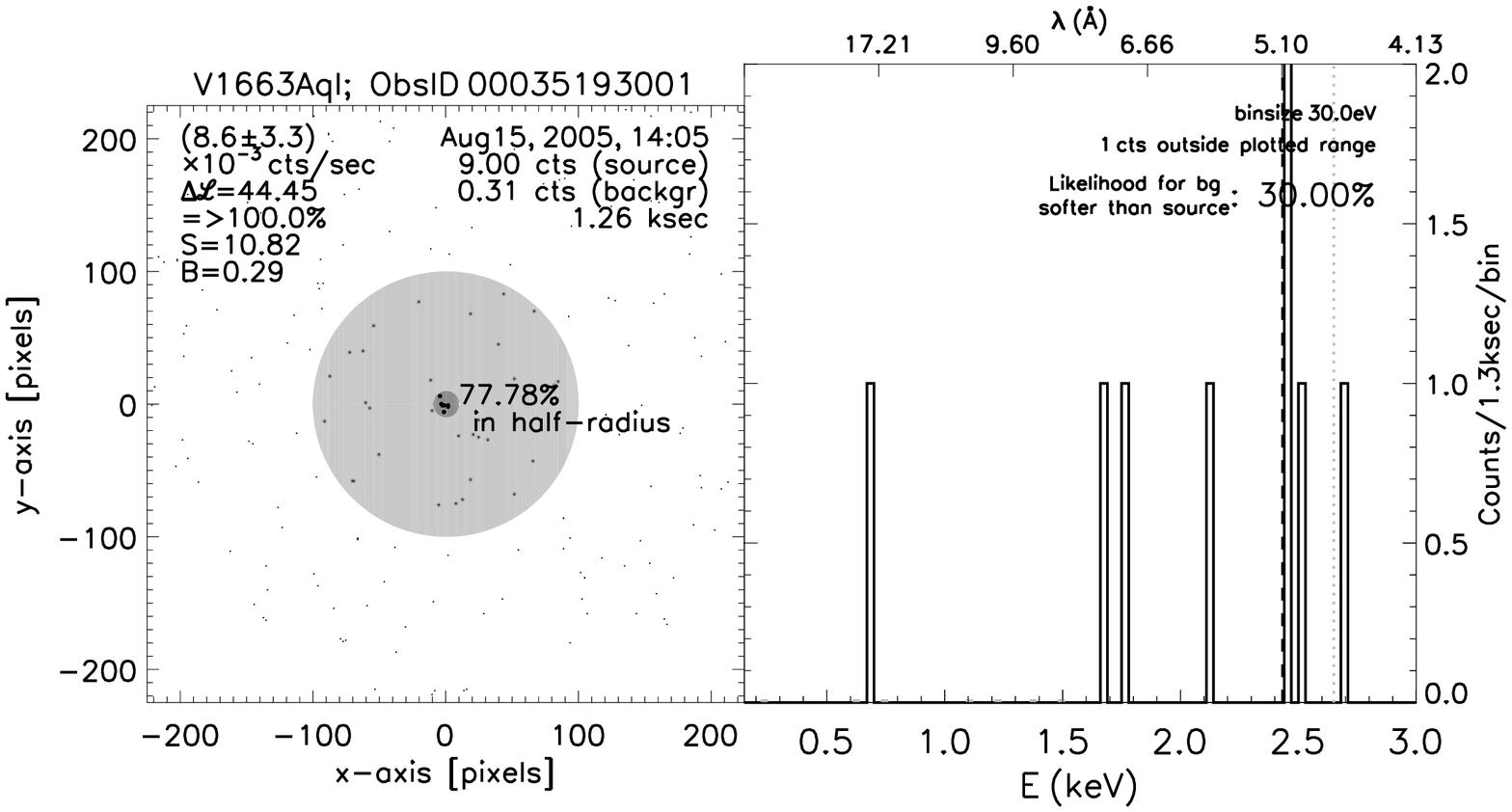}}

\resizebox{\hsize}{!}{\includegraphics{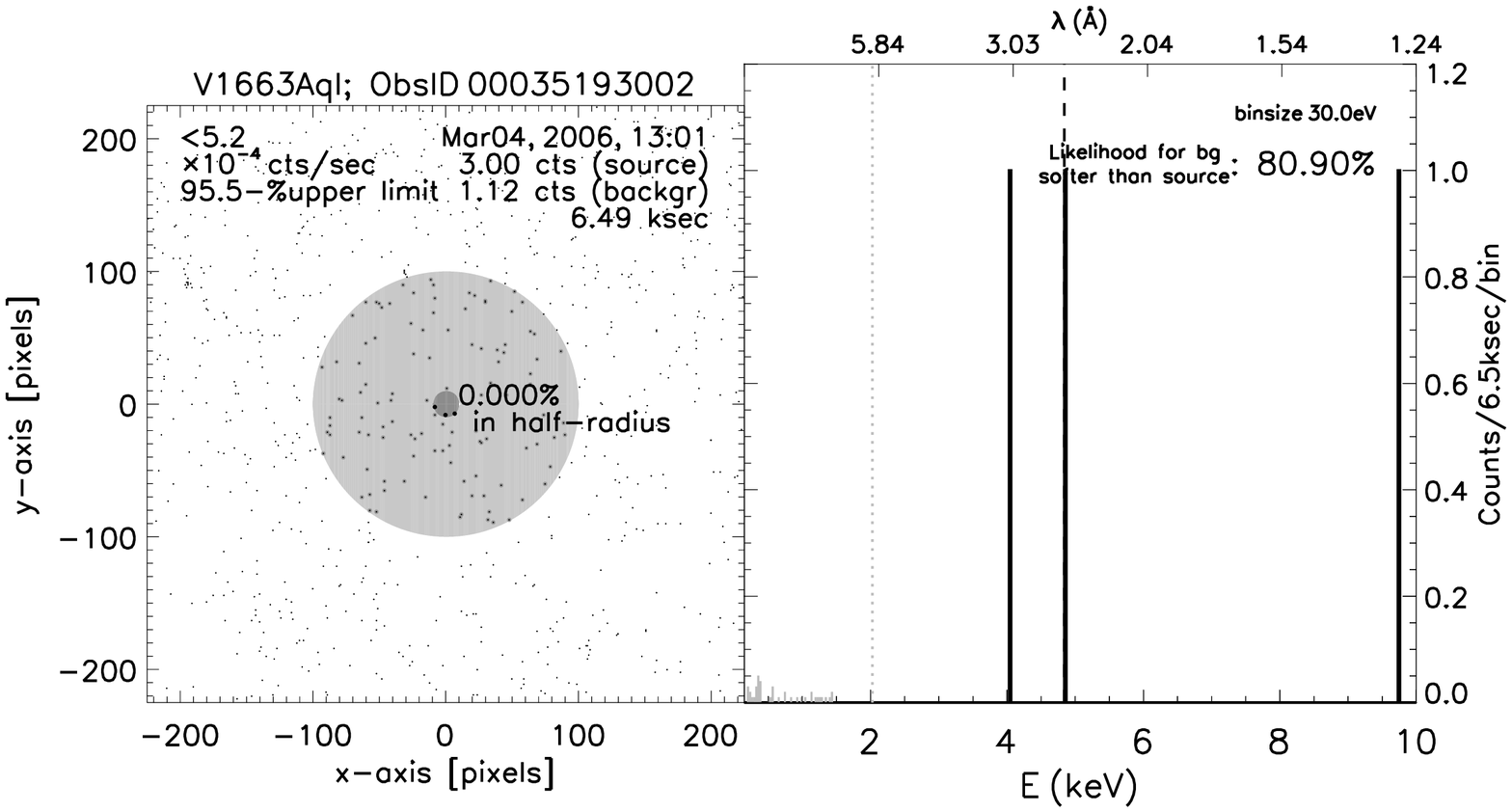}}
\caption{\label{V1663aql}Two \swift\ observations of V1663\,Aql.
The left panels (image) are the same as described in
Fig.~\ref{firstphot}. The right panels provide the spectral information of
the source (black) and background (downscaled to the area of the source extraction
region; light shadings). The likelihood for the
background spectrum being softer than the source spectrum is estimated by
drawing 1000 random sample spectra with the number of source counts (in the
upper panel 9) out of the background ($0.31\times 99=31$ counts) and
counted the number of cases where the median energy of each sample
spectrum was lower than that of the source spectrum. Much different values
than 50\,percent are a soft criterion for a source detection. In the
upper panel we find a clear detection and give the count rate with 1-$\sigma$
uncertainty, $\Delta {\cal L}$ according to Eq.~\ref{deltal}, and the number of
net counts for source and background $S$ and $B$ (per detect cell) according to
Eq.~\ref{model}.}
\end{figure*}

V1663\,Aql was the first nova observed by \swift. Unfortunately, the first
planned observation was truncated by a GRB after only 1.25\,ksec.
Nevertheless, we find
a clear detection and a strong concentration of
counts towards the center (Table~\ref{obstab}). The source
spectrum is similar
to the background spectrum, but that does not rule out
a detection (see Fig.~\ref{V1663aql}). The spectrum is extremely
weak and
no conclusion can be drawn. A second observation seven months later was
longer but yielded only an upper limit which implies a fading of
the source. Neither of our other two criteria supported a detection.

\begin{figure*}
\resizebox{\hsize}{!}{\includegraphics{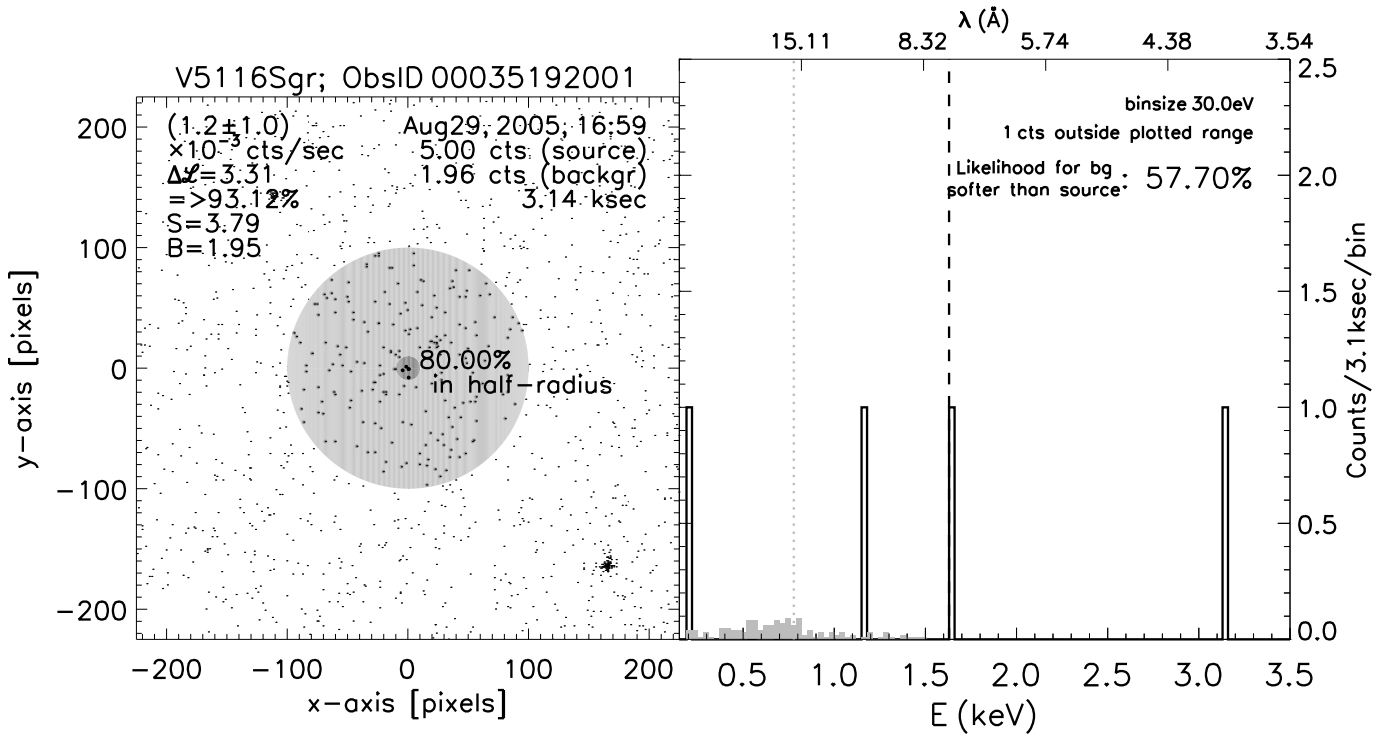}}
\caption{\label{V5116sgr}Same as Fig.~\ref{V1663aql} but for V5116\,Sgr.}
\end{figure*}

For V5116\,Sgr we only found a marginal (formal 93-percent) detection
with a concentration of photons towards the center. The spectrum
is too weak to apply our source-background comparison criterion
(see Fig.~\ref{V5116sgr}).

\begin{figure*}
\resizebox{\hsize}{!}{\includegraphics{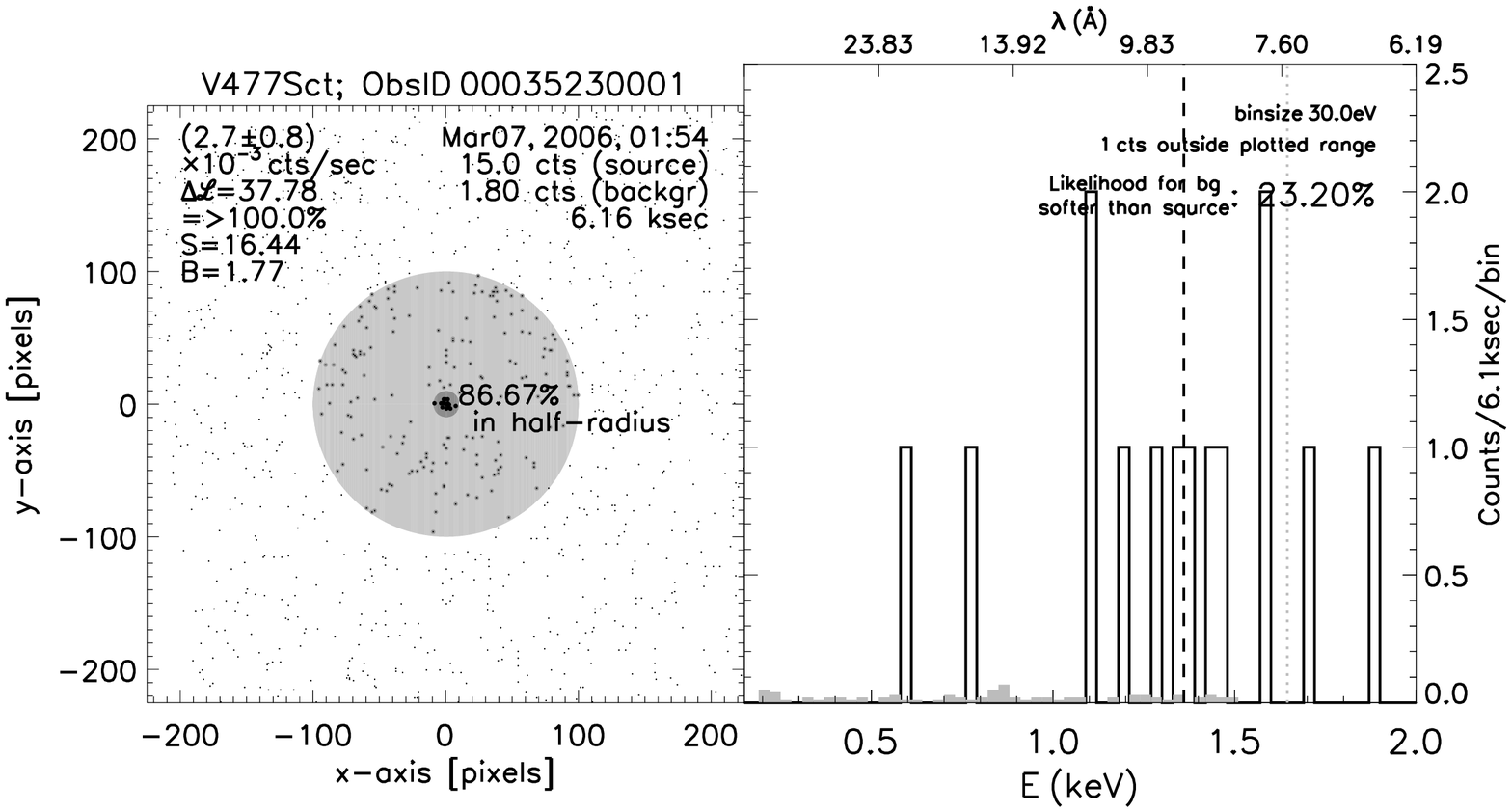}}
\resizebox{\hsize}{!}{\includegraphics{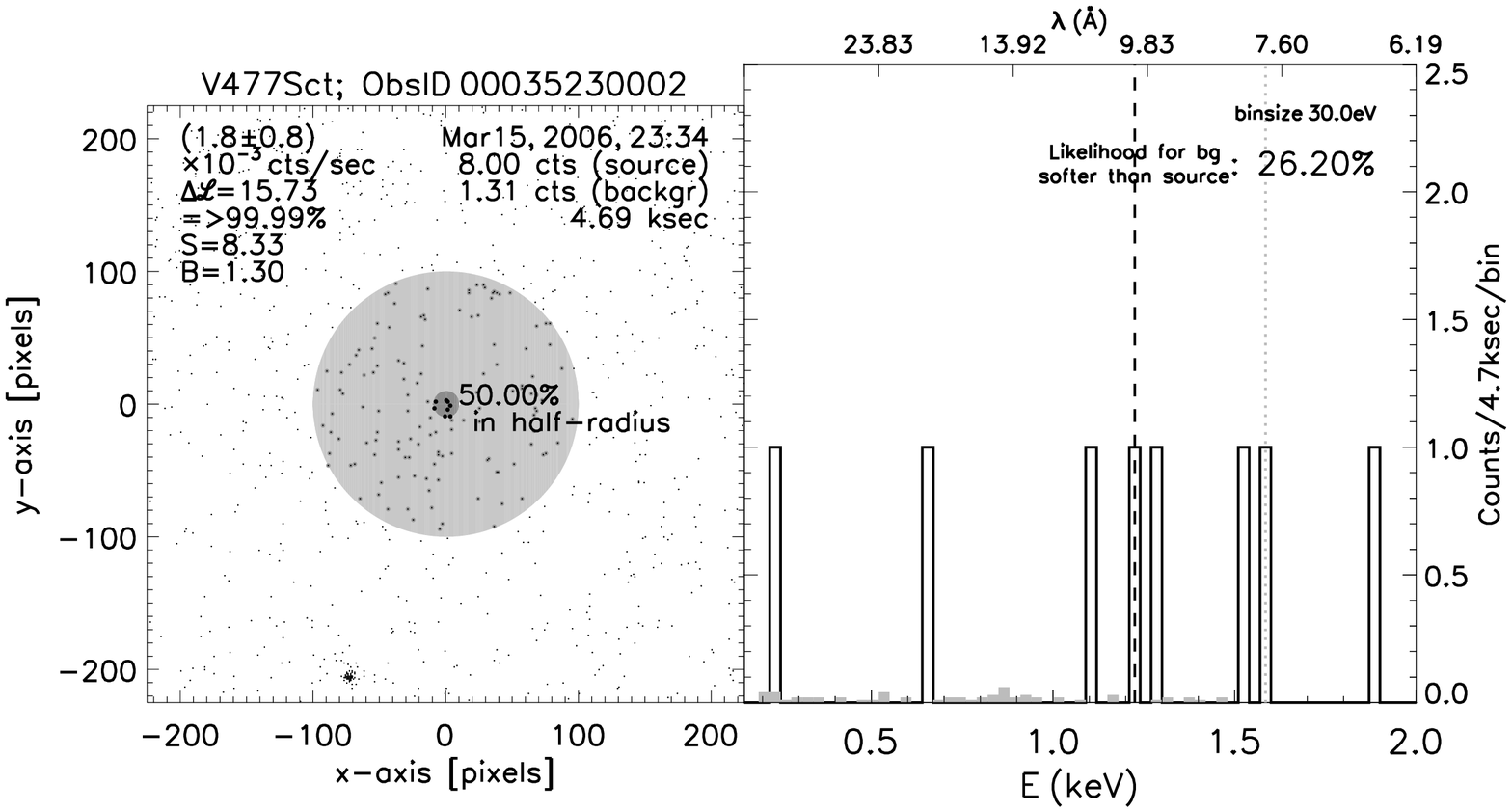}}
\caption{\label{V477sct}Same as Fig.~\ref{V1663aql} but for the two
observations V477\,Sct.}
\end{figure*}

Two observations of V477\,Sct were carried out 8 days apart and resulted in
two detections with a likelihood $>99$\,percent (see
Fig.~\ref{V477sct}). The first
observation suggests a weak source with a clear concentration towards
the center and a much softer spectrum than that of the instrumental background.
The second observation reveals a lower count rate, lower detection
likelihood, and a less clear concentration of photons towards the
center. While this implies variability or that the source has faded over the
eight day interval, within
the 1-$\sigma$ errors, the count rate is consistent with that of the first
observation. Neither spectrum allows quantitative analysis.

\begin{figure*}
\resizebox{\hsize}{!}{\includegraphics{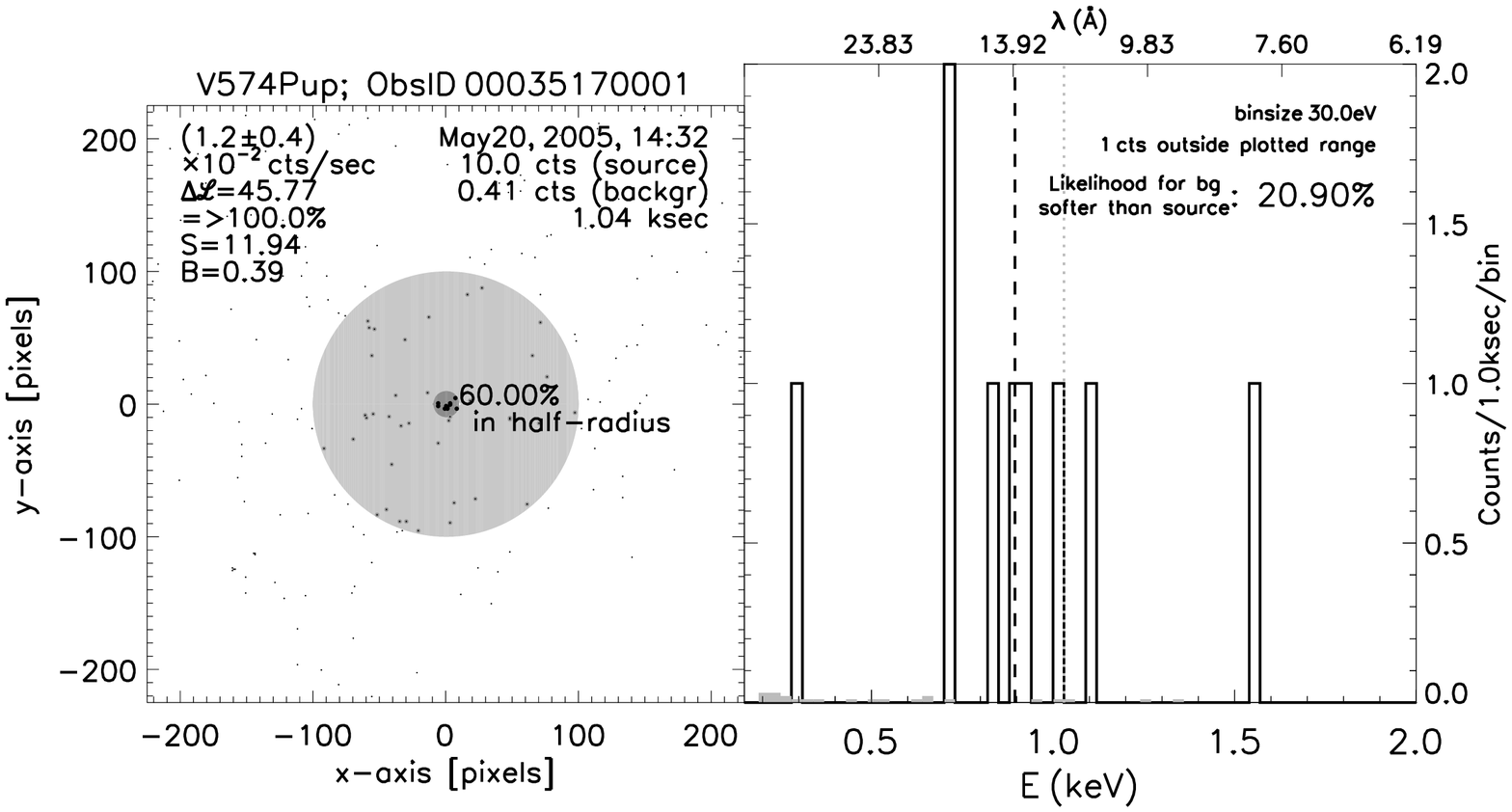}\includegraphics{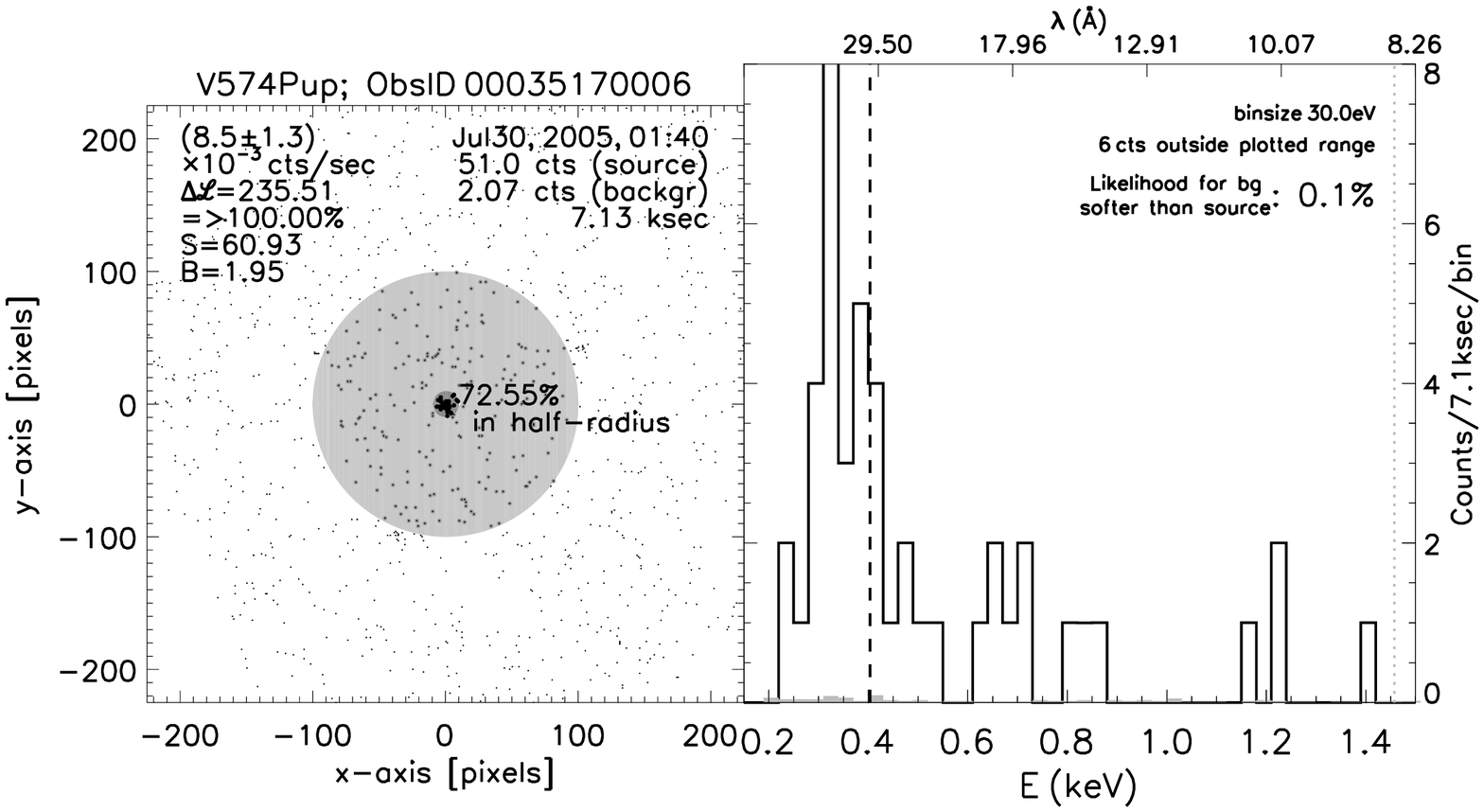}}
\resizebox{\hsize}{!}{\includegraphics{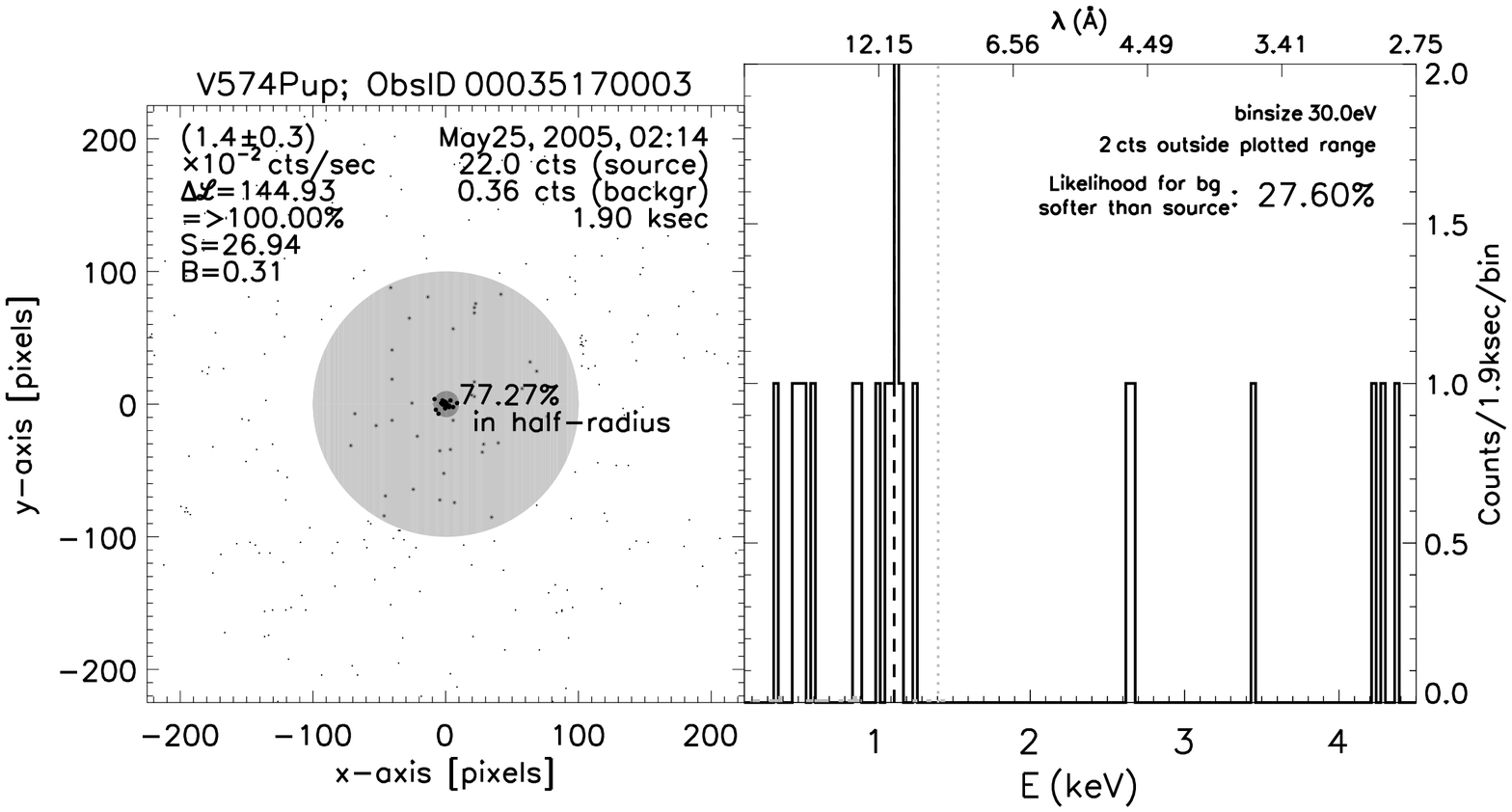}\includegraphics{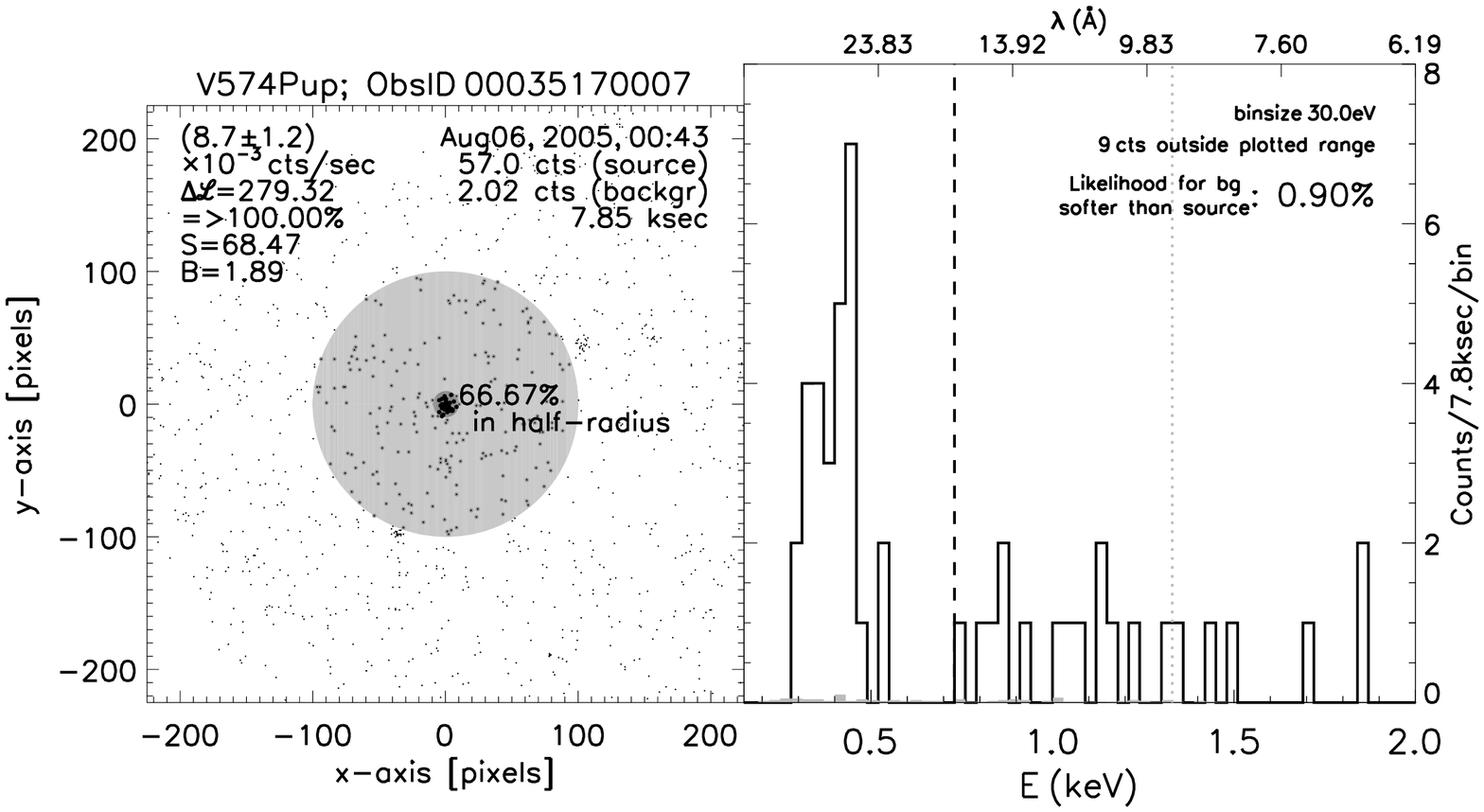}}
\resizebox{\hsize}{!}{\includegraphics{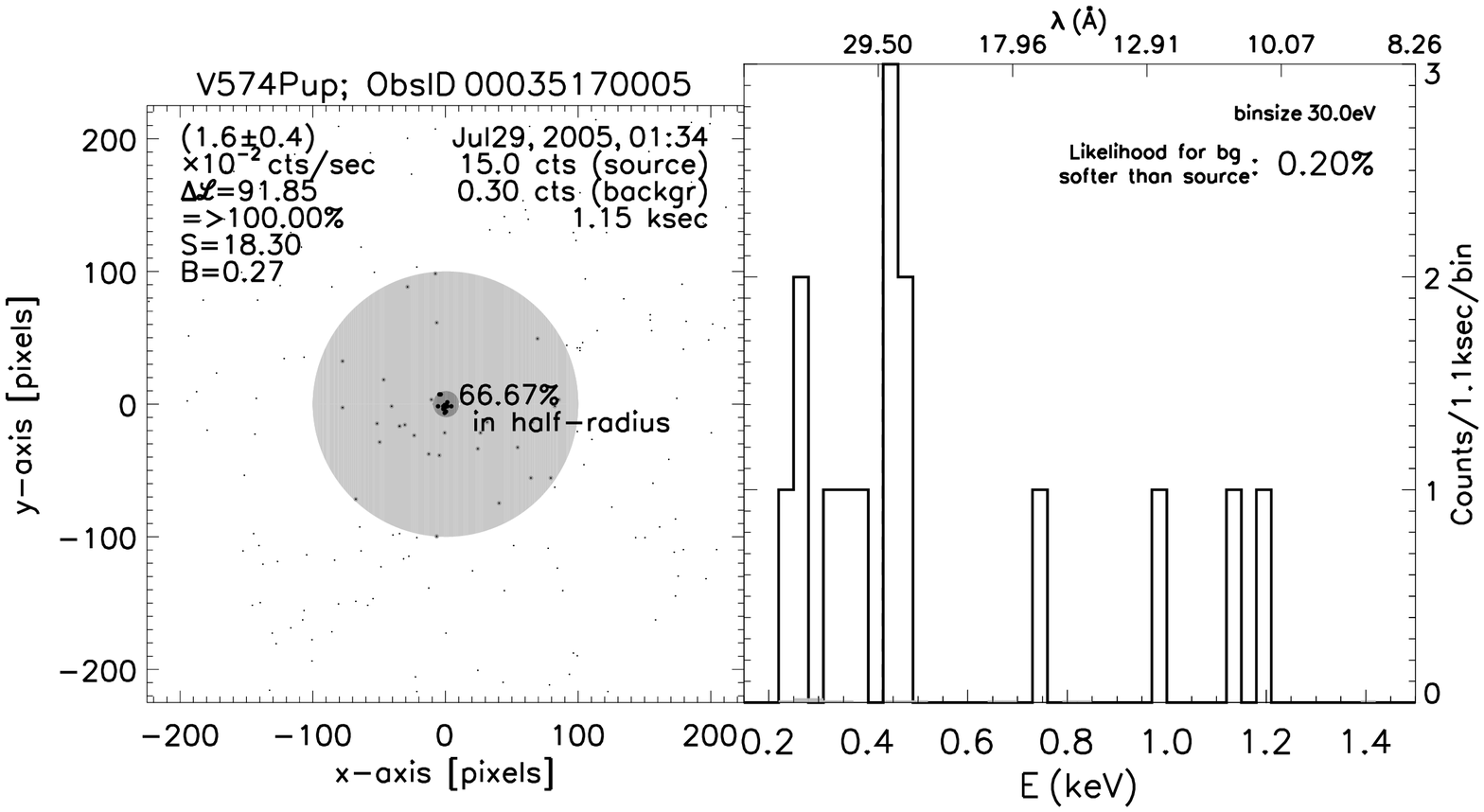}\includegraphics{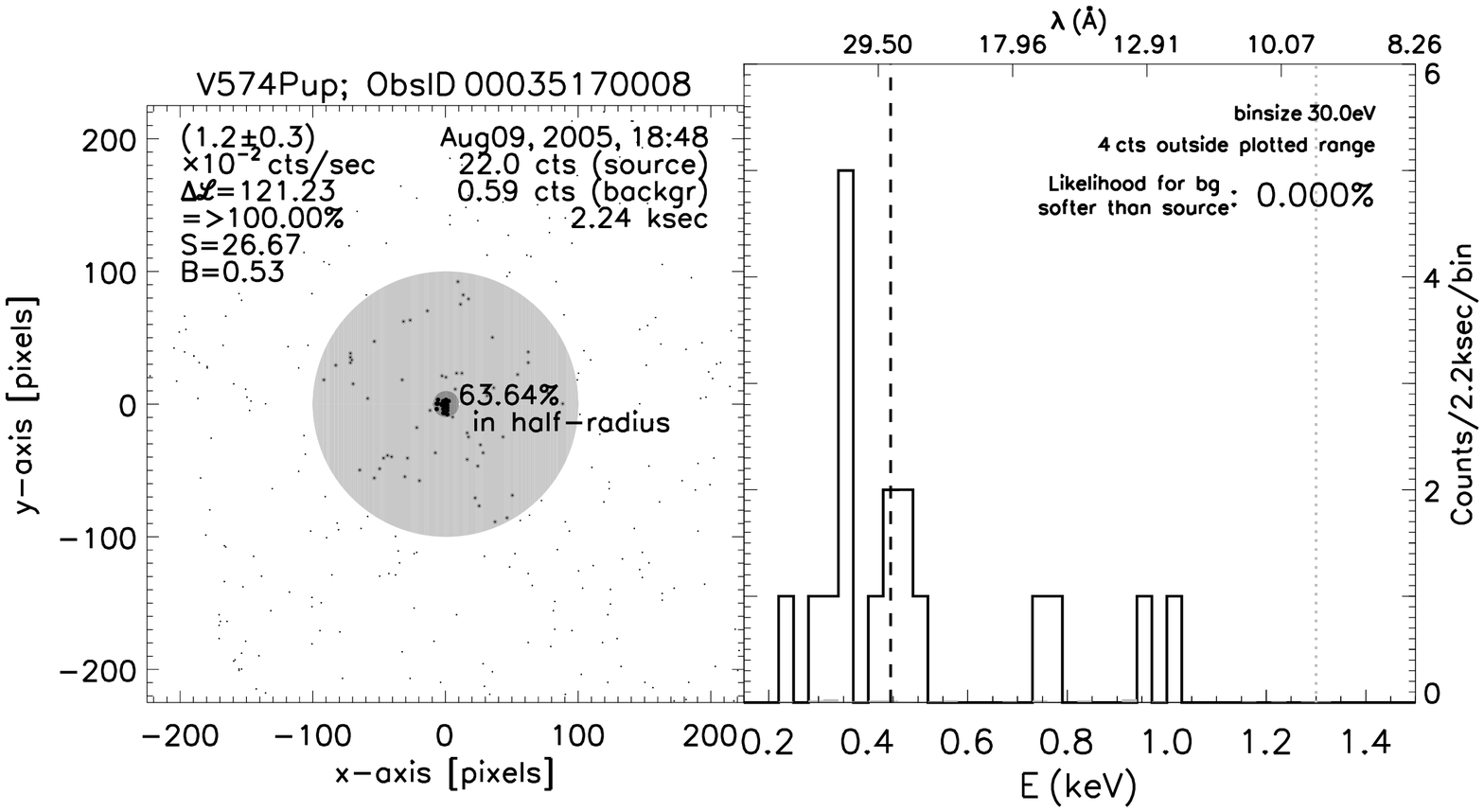}}
\vspace{.2cm}

\caption{\label{V574pupall}Same as Fig.~\ref{V1663aql} but for
six usable observations of V574\,Pup.}
\end{figure*}


\begin{figure*}
\resizebox{\hsize}{!}{\includegraphics{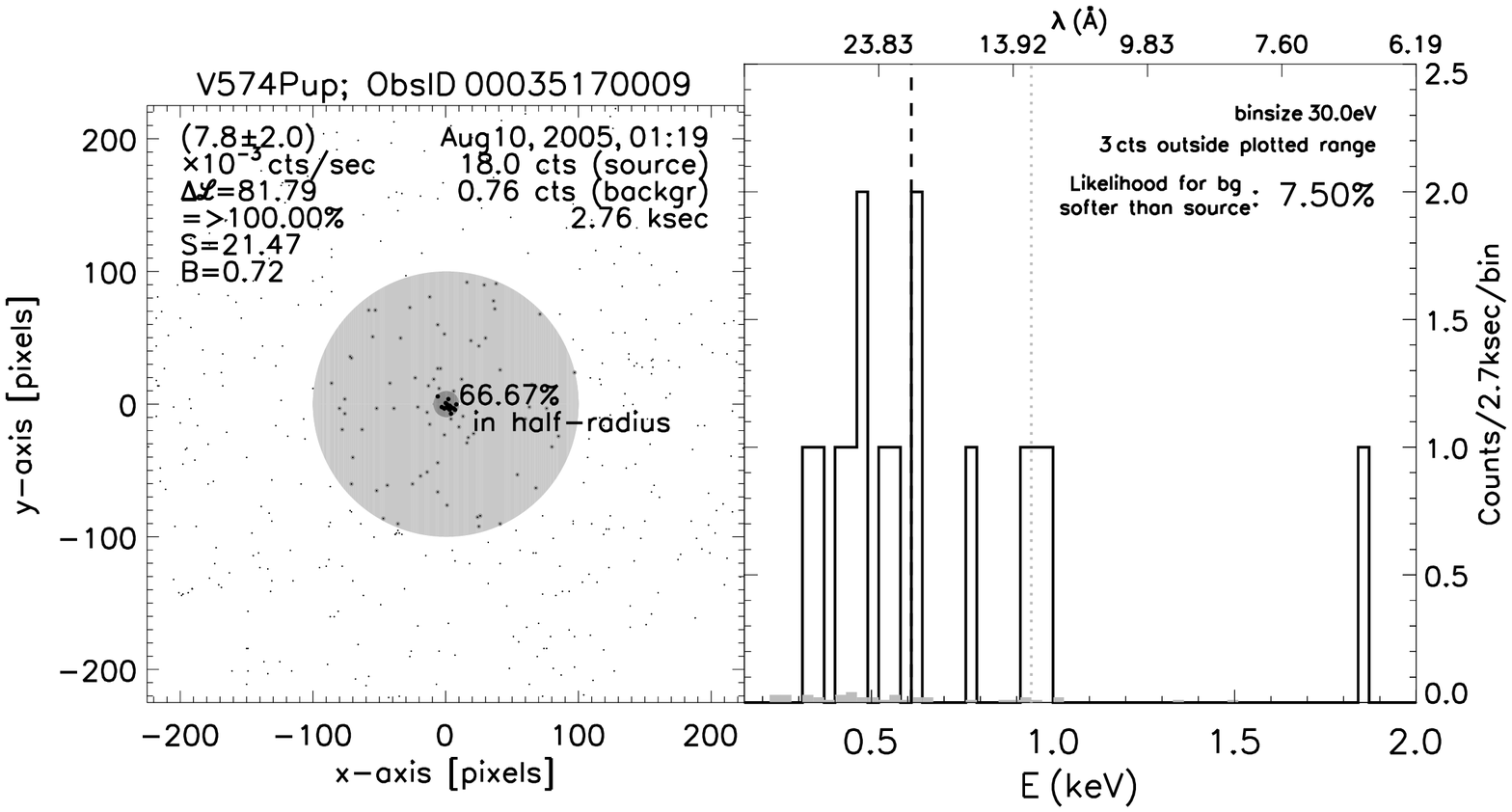}\includegraphics{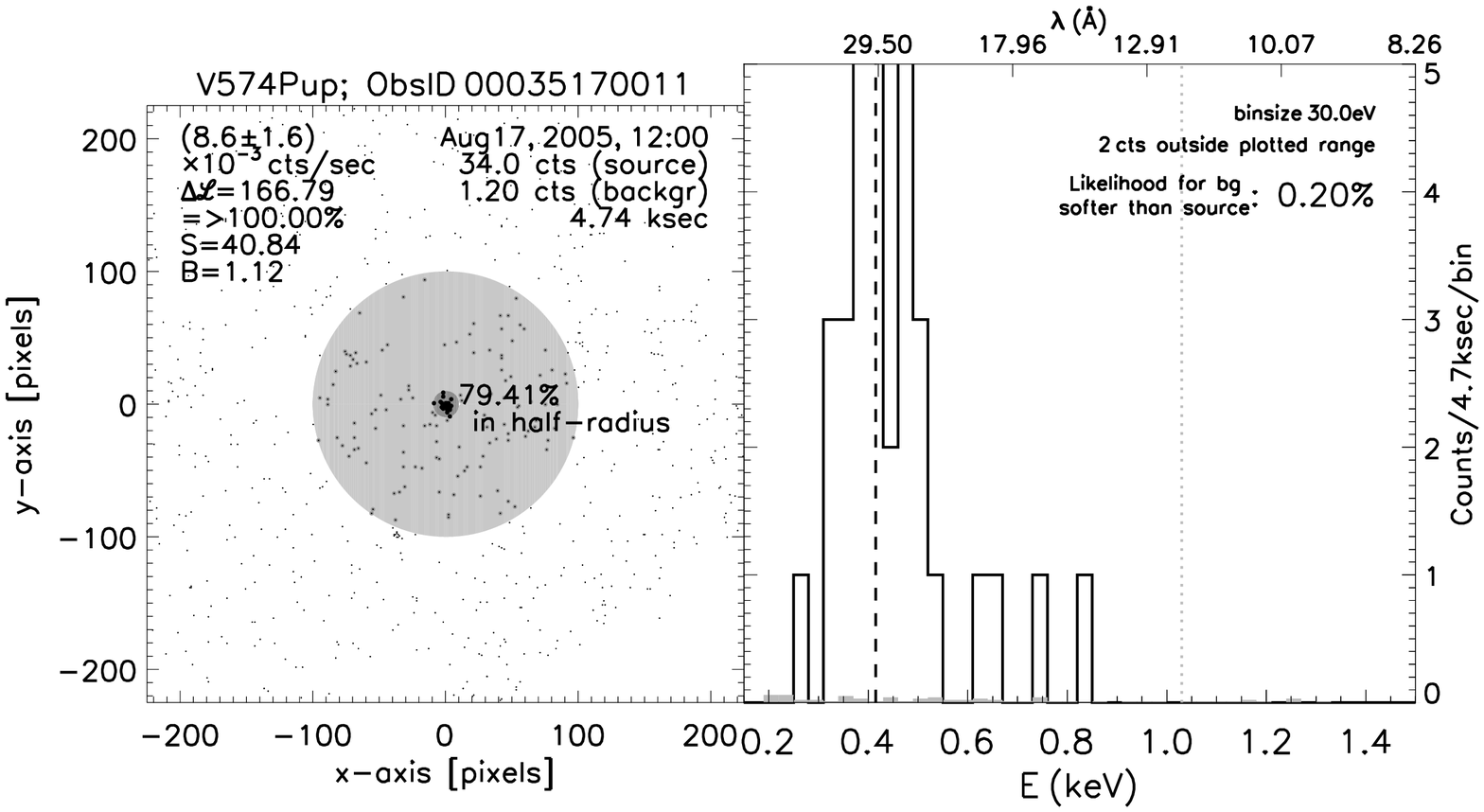}}
\resizebox{0.5\hsize}{!}{\includegraphics{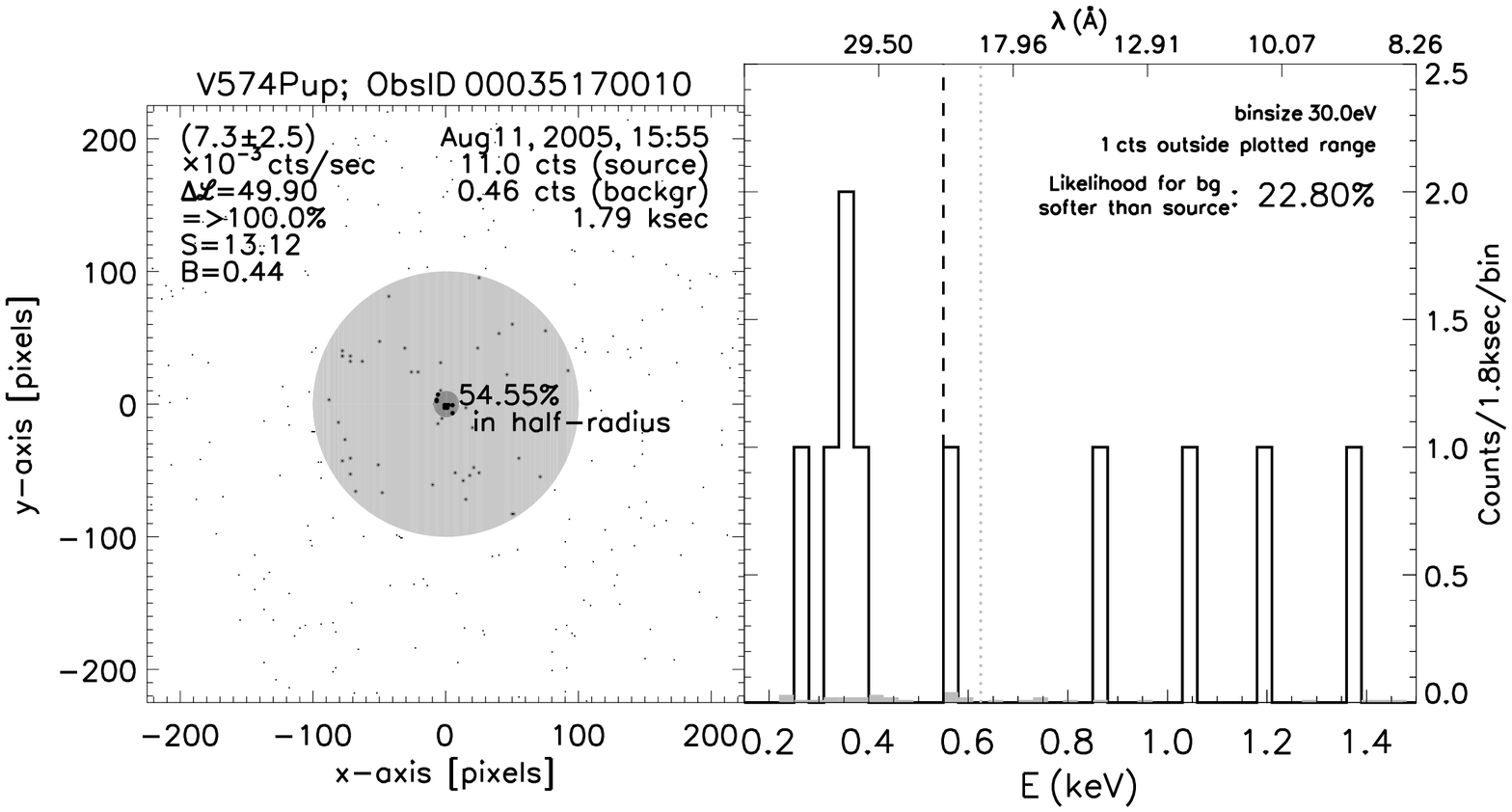}}
\vspace{.2cm}

\caption{\label{V574pupall2}Continuation of Fig.~\ref{V574pupall}.}
\end{figure*}


\begin{figure*}
\resizebox{\hsize}{!}{\includegraphics{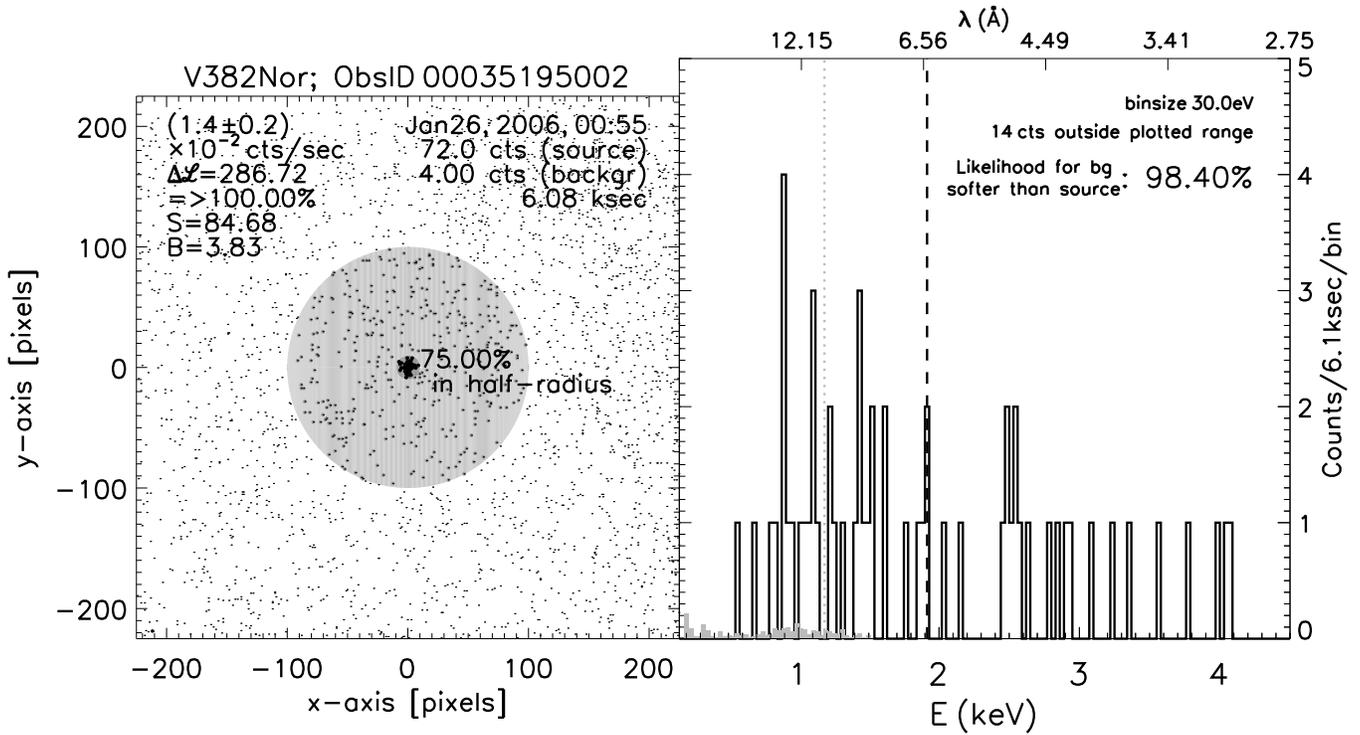}}
\caption{\label{V382nor}Same as Fig.~\ref{V1663aql} but for V382\,Nor. The
spectrum appears to be an emission line spectrum, but we note that it is oversampled
and the lines cannot be as clearly identified.}
\end{figure*}


\begin{figure*}
\resizebox{\hsize}{!}{\includegraphics{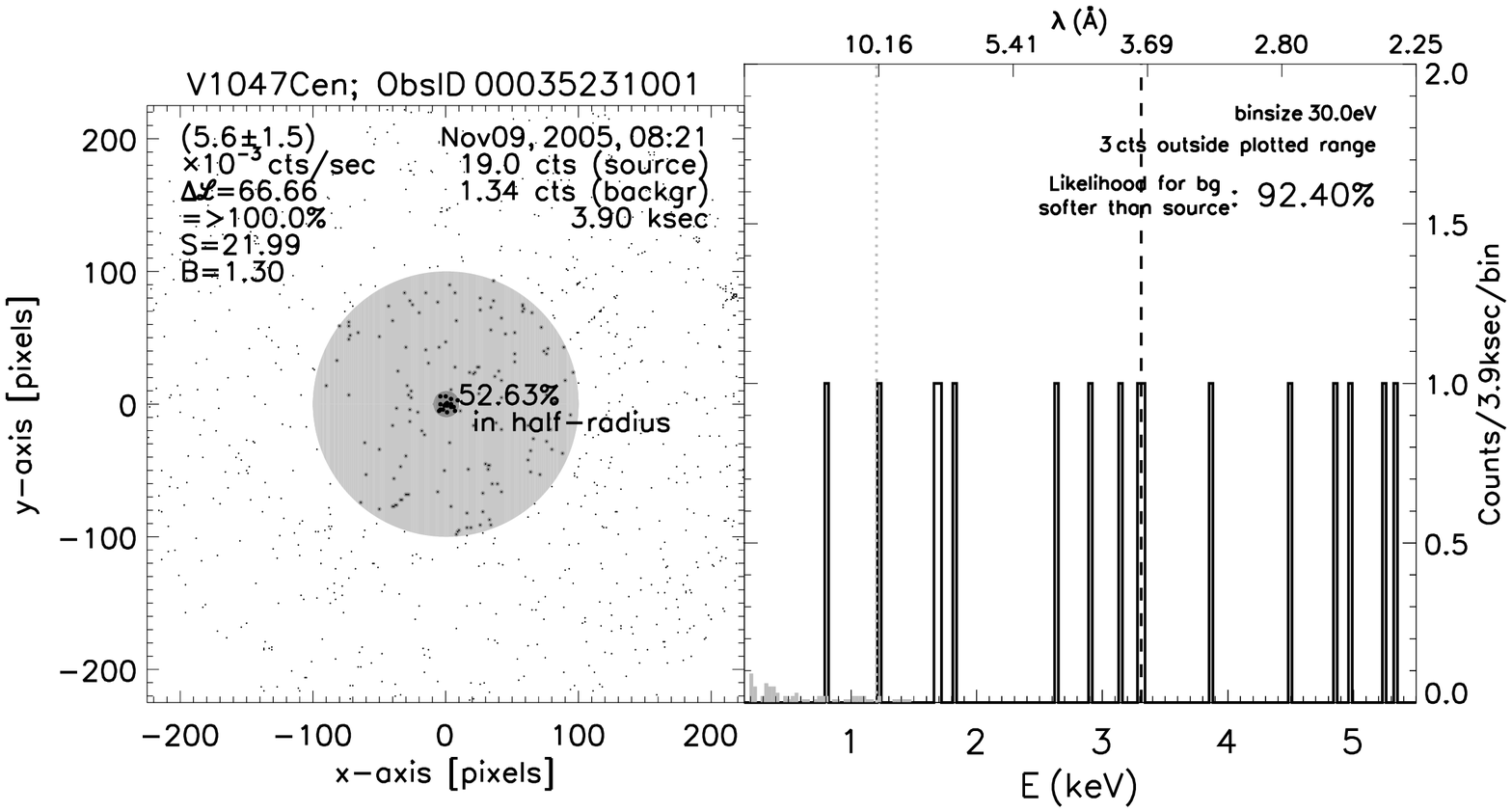}}
\resizebox{\hsize}{!}{\includegraphics{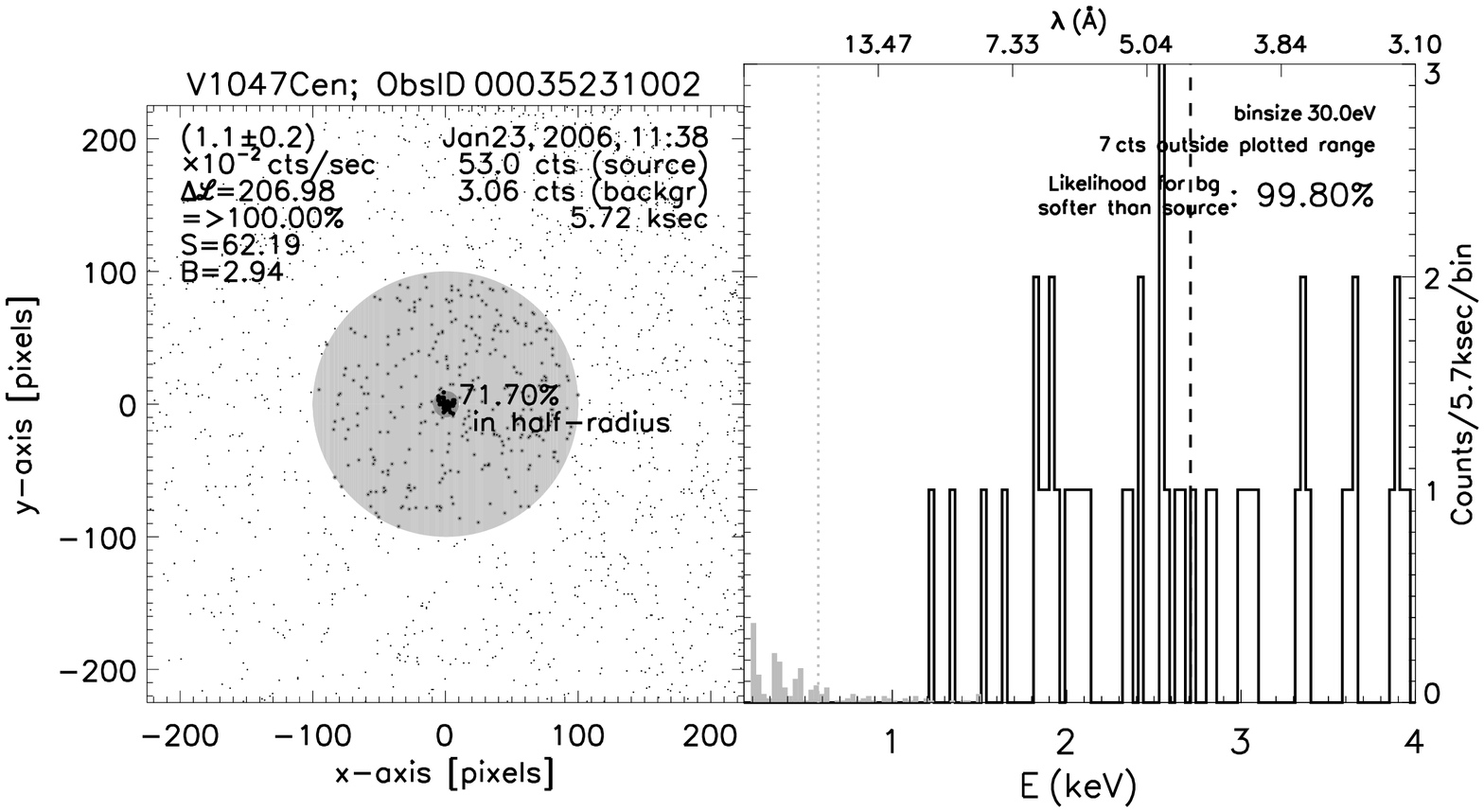}}
\caption{\label{V1047cen}Same as Fig.~\ref{V1663aql} but for the two
observations of V1047\,Cen.}
\end{figure*}

\begin{figure*}
\resizebox{\hsize}{!}{\includegraphics{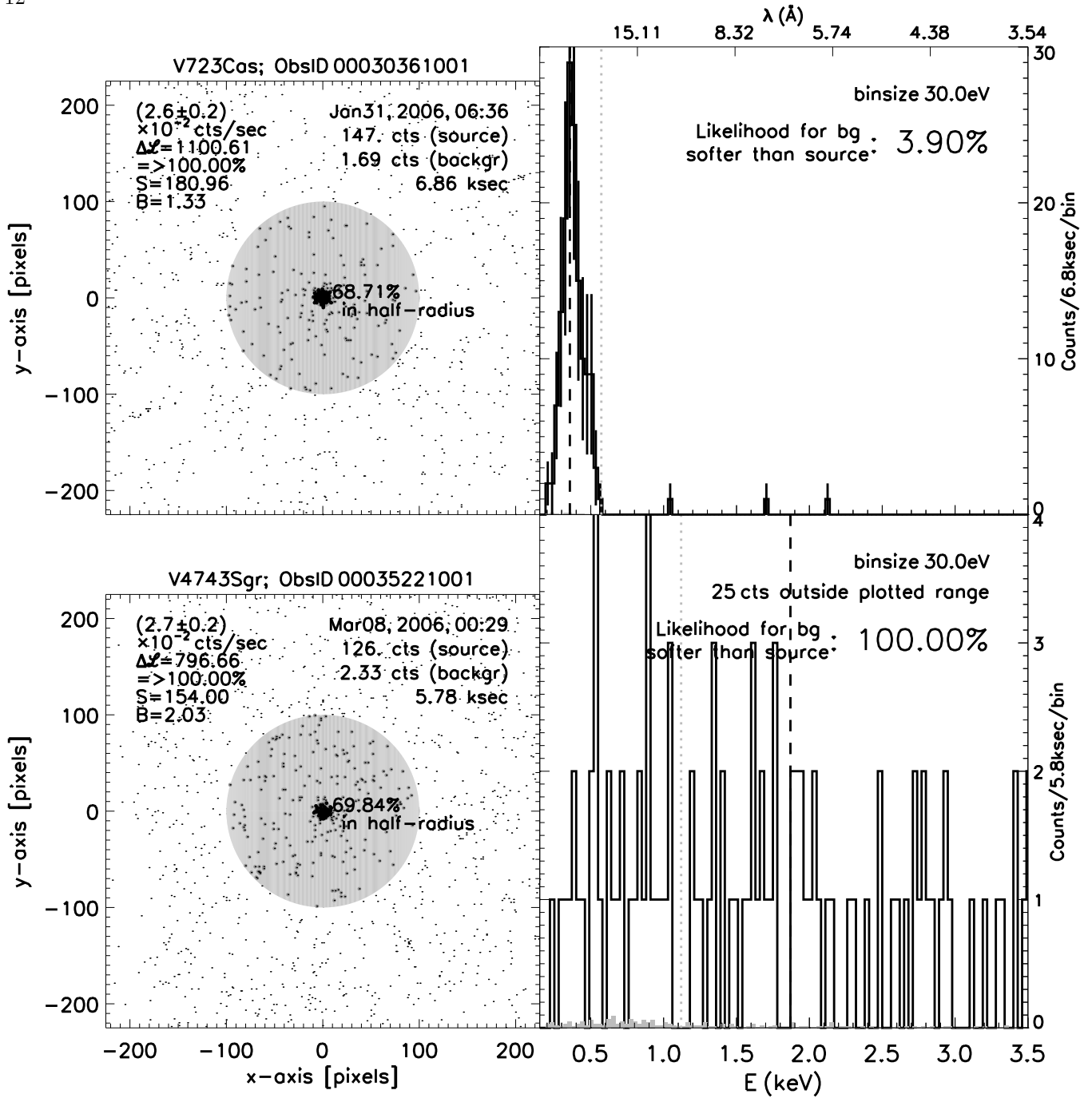}}
\caption{\label{lastphot} Same as Fig.~\ref{V1663aql} but for V723\,Cass (top)
and V4743\,Sgr (bottom). While the count rates are similar, the spectra
are very different.}
\end{figure*}

\begin{figure}
\resizebox{\hsize}{!}{\includegraphics{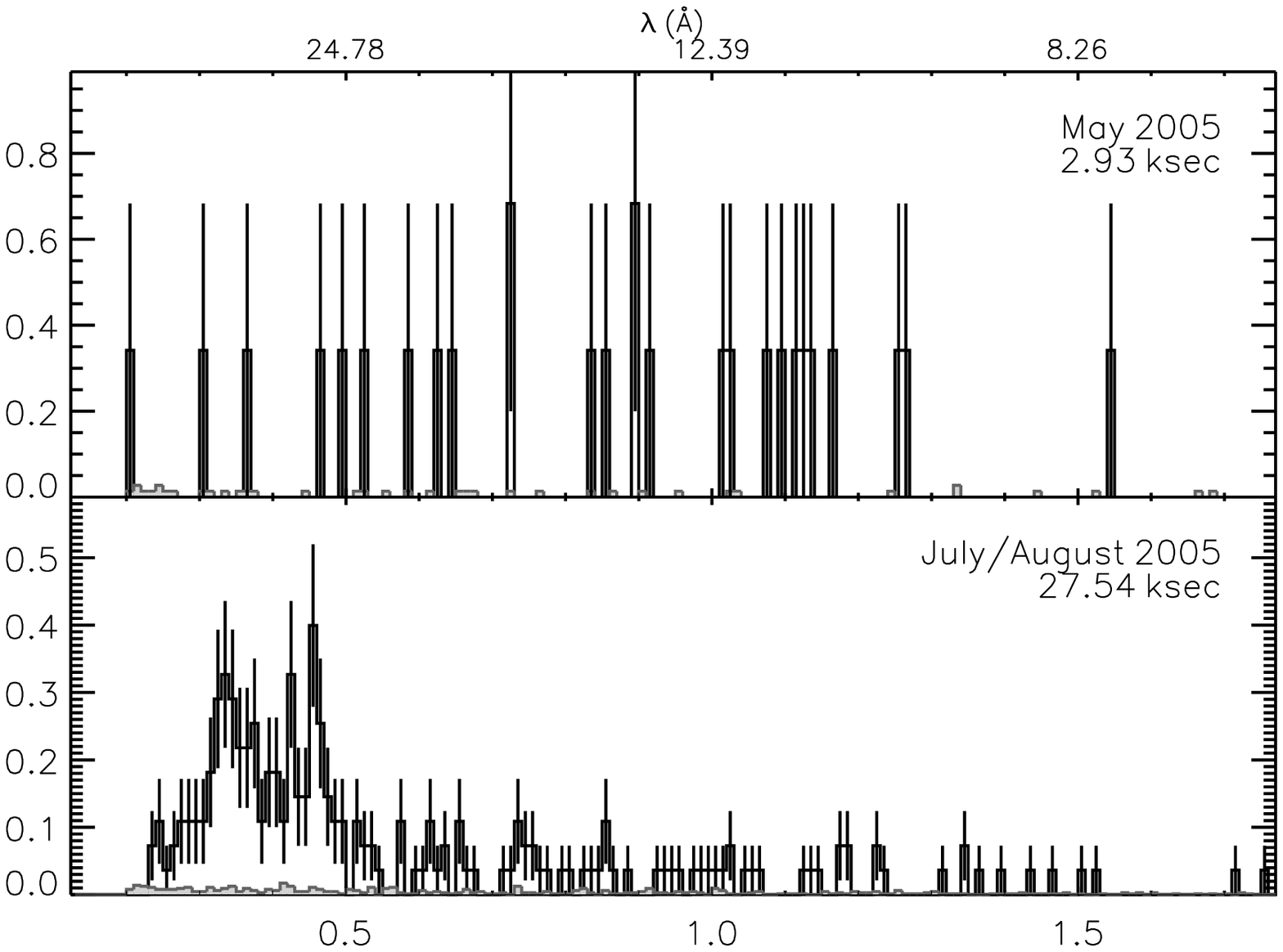}}
\caption{\label{cmp_v574}Grouped XRT observations of V574\,Pup comparing
the May 2005 (top panel) and the July through August data
sets (bottom panel). The comparison shows that V574\,Pup has evolved from a hard
early spectrum (top panel) into a SSS spectrum (bottom). The count rate is higher
in the top panel, which may be due either to the higher sensitivity of the detector at
higher energies or the higher amounts of absorption at soft energies.}
\end{figure}

\begin{figure}
\resizebox{\hsize}{!}{\includegraphics{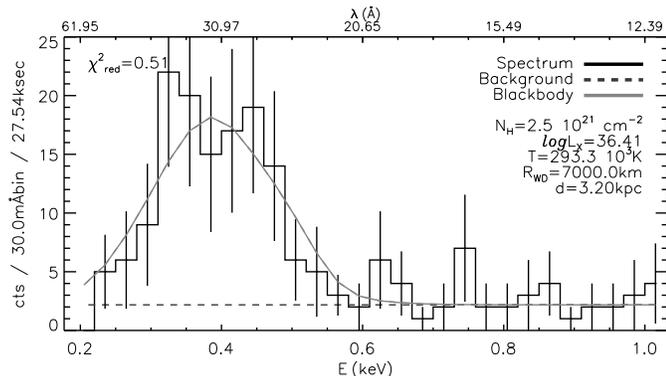}}
\caption{\label{V574Pup_bbb}Grouped and rebinned XRT spectrum of
V574\,Pup
covering from July 2005 to August 2005 (shown in the bottom
panel of Fig.~\ref{cmp_v574}). The histogram (plus error bars)
gives the observed spectrum and smooth line shows the
best-fit blackbody model
overplotted (parameters given in legend).}
\end{figure}

\begin{figure}
\resizebox{\hsize}{!}{\includegraphics{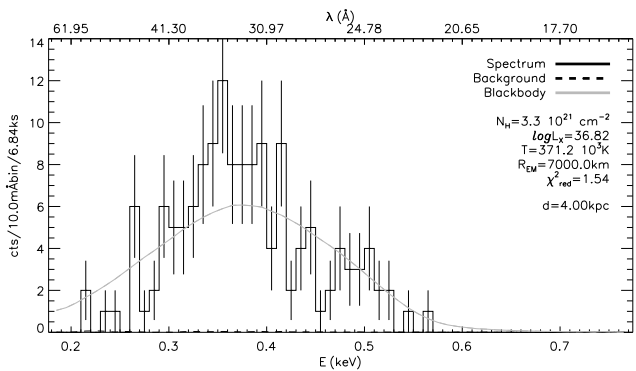}}
\caption{\label{v723cas_bb}Blackbody fit to the XRT spectrum of
V723\,Cas. The background lies on top of the horizontal axis.}
\end{figure}

\begin{figure}
\resizebox{\hsize}{!}{\includegraphics{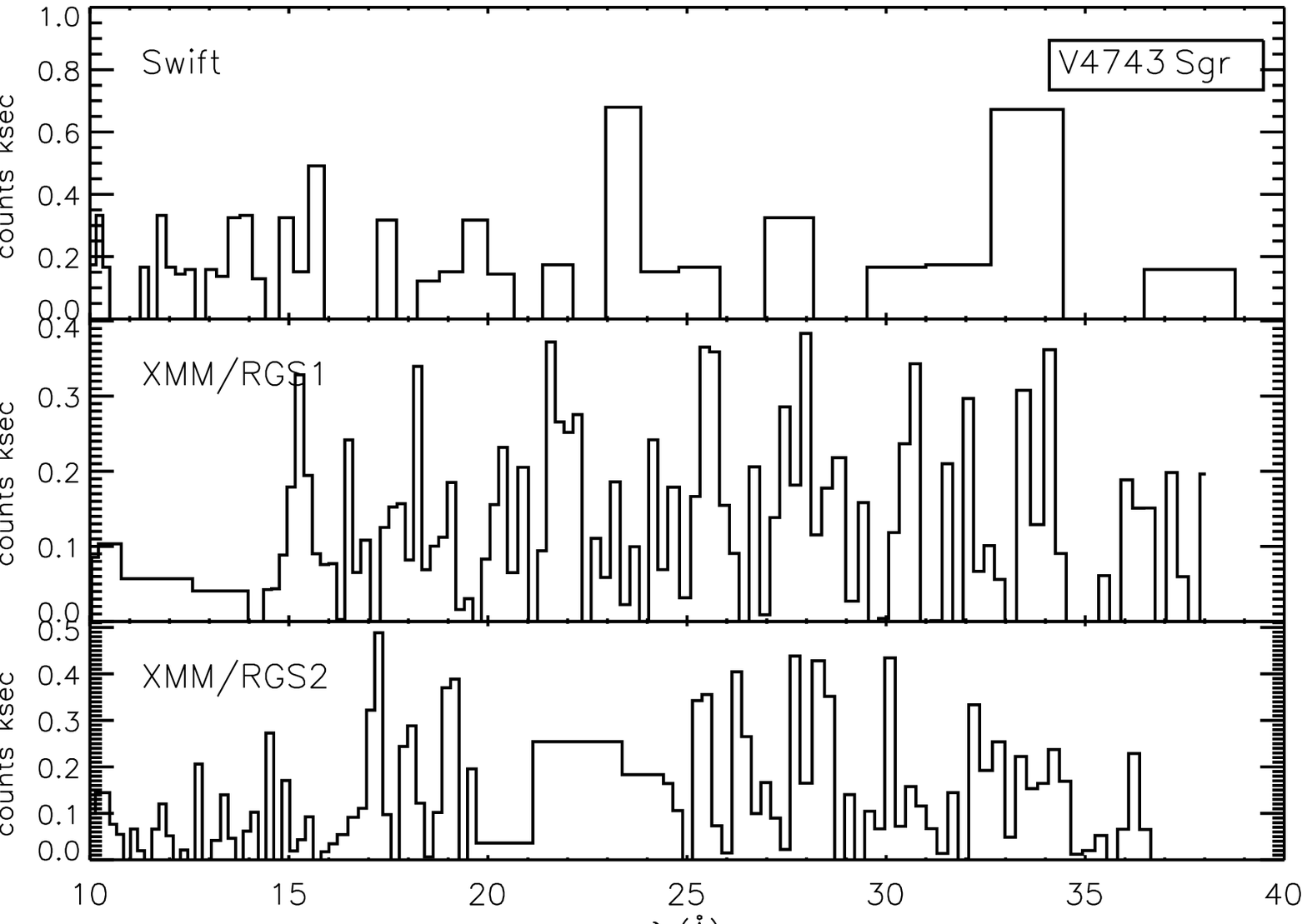}}
\caption{\label{cmp_sgr}Comparison of the XRT spectrum of V4743\,Sgr (top)
with \xmm\ RGS1 and RGS2 spectra from September 2004. All are consistent with
a fading nebular spectrum, but the signal to noise and the resolution are
insufficient to characterize the spectrum as hard continuum or an emission
line spectrum.}
\end{figure}


\subsection{Strong sources $>0.01$\,cps}
\label{clear}

V574\,Pup, V382\,Nor, V1047\,Cen, V723\,Cas, and V4743\,Sgr were
detected by \swift\ (see Figs.~\ref{V574pupall}--\ref{lastphot}).
The spectrum of V382\,Nor is hard whereas the background is
softer. There appears to be some excess emission at wavelengths of prominent
emission lines (e.g., 0.92\,keV, \ion{Ne}{9}; 1.02\,keV, \ion{Ne}{10};
and 0.83\,keV, \ion{Fe}{17}) but a firm identification of these
lines is not
possible with the existing data. There is also emission
at energies with no well-known emission lines, particularly at high energies.
The observed spectrum is too faint to derive any quantitative conclusions.

We obtained very clear detections of V1047\,Cen. In the first
observation of November 2005 the photons are clearly concentrated towards the
center, and the energy distribution is very different
from the soft background. Two months
later the detection is even stronger and the spatial as well as spectral
distributions clearly support a detection. The count rate doubled from
5.6\,cts\,ksec$^{-1}$ on 9 November, 2005 to 11\,cts\,ksec$^{-1}$ on
23 January, 2006. The spectrum appears hard, but there are too
few counts to derive any spectral properties.

The remaining novae were sufficiently bright to obtain spectral
confirmation that they were
detected during the SSS phase. More detailed analysis of these spectra,
including additional \swift\ observations and supporting optical and
near-IR spectra, is left for future papers. Here we only provide a brief
analysis of the current \swift\ data sets.

V574\,Pup was first observed in late May 2005. It was then observed
multiple times from the end of July to mid August 2005 as it was perfectly
positioned to cool the XRT detectors between GRB observations. In all nine
observations, it was clearly detected with a high degree of significance
(see Table~\ref{obstab}).
V574\,Pup had an average of $7-14$\,cts\,ksec$^{-1}$ in each observation.
The individual observations were too short for analysis, so we combined them,
yielding a higher signal-to-noise ratio. We grouped the spectra
by summing the number of counts and the number of background counts
in each spectral bin. We calculated exposure time-weighted effective areas
from the individual extracted effective areas.

 However, during the 50 days between the May 26 and July 29 observations,
significant evolution probably occurred which we would miss by combining all
spectra together. In order to search for any evolution, we grouped the first
two observations from May 2005 and the remaining observations from July and
August 2005. These are shown in Fig.~\ref{cmp_v574}. The summed
spectra of the two early
observations shows a uniform distribution with energy and may be an underexposed
emission line spectrum. Unfortunately, we don't have any more data from this time
period to increase the signal-to-noise ratio. The later observations
sum up to what appears to be a continuous spectrum ranging from 0.3--0.5\,keV
(25\,\AA\ to 45\,\AA). This spectrum resembles that of a SSS. We carried out a
blackbody fit using the most recent XRT response matrices (version 8)
and assuming an emitting radius of 7000\,km. We found a satisfactory fit
(Fig.~\ref{V574Pup_bbb}) with the parameters
T$_{\rm eff} = (293\,\pm\,10)\times10^3$\,K ($25.2\,\pm\,1$\,eV)
and N$_{\rm H} = (2.5\,\pm\,0.6)\times10^{21}$\,cm$^{-2}$ (1-$\sigma$ uncertainty ranges). We
fixed the radius because no unique solution could be found with a variable
radius. The uncertainties include variations of the radius
between 7000 and 9000\,km.

The XRT spectrum of V4743\,Sgr is faint, and no spectral features can be
identified. We also found no features in the RGS observation taken in
September 2004 (see below). While the XRT count rates of V4743\,Sgr
and V723\,Cas are similar, the spectra are different
(Fig.~\ref{lastphot}). V723\,Cas shows a clear peak occurring at
0.4\,keV (31\,\AA).
\citet{2006IAUC.8676....2N} reported that the spectrum resembled a SSS, and they
were able to fit a blackbody model. We repeated these fits using
the updated XRT response matrices
(version 8) and show the best-fit model in Fig.~\ref{v723cas_bb}.
We fixed the radius at 7000\,km but included
variations up to 9000\,km in the calculation
of the 1-$\sigma$ uncertainty ranges. We find
T$_{\rm eff} = (371\,\pm\,15)\times10^3$\,K ($32\,\pm\,2$\,eV)
and N$_{\rm H} = (3.3\,\pm\,0.4)\times10^{21}$\,cm$^{-2}$. V723\,Cas is now, after 11
years, the oldest Galactic nova that is still active. Such a
long duration of
the SSS phase is unusual for a Classical Nova, and there is a possibility
that the nova has transitioned into a SSS such as Cal\,83. Our observations of this interesting source are
continuing.

\subsection{Supplemental observations}
\label{supext}

 In addition to the \swift\ observations, we extracted unpublished
X-ray observations from the \chandra\ and \xmm\ archives.
 A series of \chandra\ observations of V1494\,Aql were carried out
and we extracted the last observation (LETGS, ObsID
\dataset[ADS/Sa.CXO#obs/2681]{2681}) taken 727\,days after outburst
(2001, Nov. 28).
The source was only detected in the zeroth order on the HRC-S
detector, but no dispersed spectrum could be extracted.
We found 56 counts on a background of 10, corresponding to an HRC count rate of
$0.02\,\pm\,0.003$\,cts\,sec$^{-1}$ and a 14-$\sigma$ detection.
Unfortunately, the HRC detector does not provide the energy resolution to
construct a spectrum, and no spectrum can be constructed from the dispersed
photons as the background is too high.

 We further analyzed \xmm\ observations of LMC\,2005 taken 243 days after
outburst (ObsID 0311591201, 2006, July 18) and of V4743\,Sgr 1470 days
after outburst (ObsID 0204690101, 2004, Sept. 30). LMC\,2005 was again not
detected.

From MOS1 we extracted $0.11\,\pm\,0.002$\,cts\,sec$^{-1}$ for
V4743\,Sgr. The source was also detected in the RGS with a count rate
of $0.01\,\pm\,0.0005$\,cts\,sec$^{-1}$. Grating spectra allow the calculation
of a flux without the need of a model, and we measured a flux of
$3\times10^{-13}$\,erg\,cm$^{-2}$\,sec$^{-1}$ from the RGS1 and
RGS2\footnote{0.35--1.8\,keV; corrected for the chip gaps}.
 Fig.~\ref{cmp_sgr} shows a comparison of these last
\xmm\ spectra with the recent \swift\ observation.

\section{Discussion}
\label{discuss}

\begin{figure*}
\resizebox{\hsize}{!}{\includegraphics{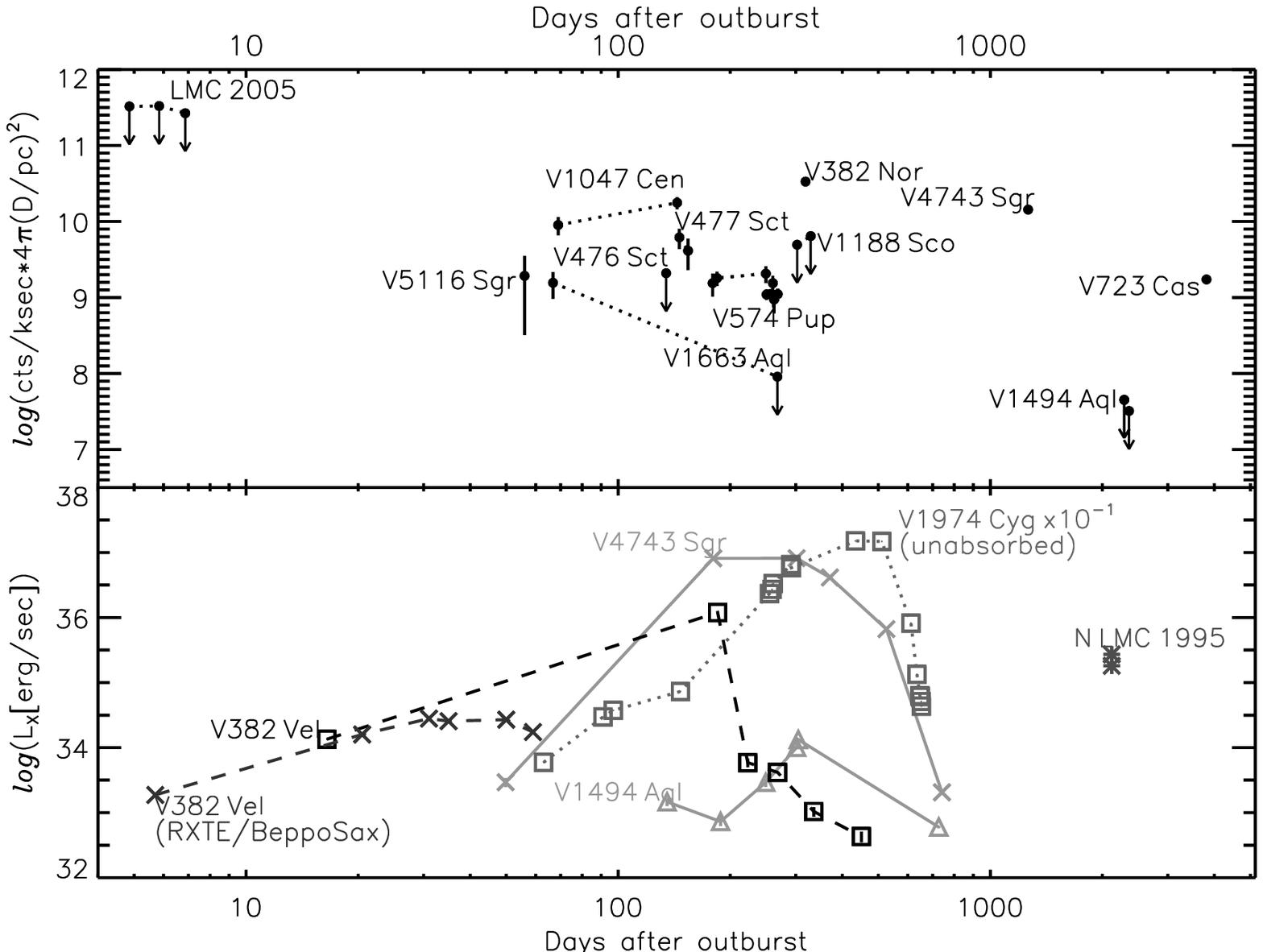}}
\caption{\label{oday}{\bf Top}: XRT count rates of all sources in our sample,
plotted against the day of observation after their outbursts. The count
rates are corrected for distance squared which is uncertain. Multiple
observations of the same targets are connected with dotted lines.
{\bf Bottom}: Observations from various other missions. The observed X-ray
luminosities (not corrected for absorption,
except for V1974\,Cyg, which is instead rescaled) of recent well-observed novae
with SSS phases. Similar plots have been created by \cite{pietsch05,pietsch06}.}
\end{figure*}

The sample of X-ray observations of novae presented in this paper demonstrates
that \swift\ significantly increased the number of Galactic novae observed in X-rays.
We provide a large sample taken with the same instrument, which reduces
problems from cross calibration. Also, \swift\ is capable of carrying out
monitoring observations with relatively little effort, the exception
being classical novae with large column densities. In Sect.~\ref{nh} we
demonstrated that one can not obtain enough counts in a typical ToO exposure
(~3-5 ks) when the column density exceeds N$_H >$ 10$^{22}$ cm$^{-2}$ to
observe an SSS phase. In this survey the estimated column densities of four
novae,
V382\,Nor, V1047\,Cen, V1663\,Aql, and V476\,Sct, exceeded this value and
only very hard spectra or non-detections were recorded.
For Galactic novae with low column densities \swift\ is an excellent
instrument for following all the stages of evolution in X-rays
\citep[e.g., the recurrent nova RS\,Oph][]{osborne06}.

 Shortly after the outburst, X-ray emission from the WD is not
expected to be observed.
X-ray detections during this phase thus originate from
other processes, e.g., shocks within the expanding shell or a shock set off
by a collision between the expanding shell and the atmosphere of the companion
\citep[as in RS\,Oph, e.g.,][]{bode06}. Few observations during this phase
have been carried out \citep[e.g.][]{mukai01} and they only provide information
on dynamics within the ejecta. With the flexibility of \swift\ a better assessment
of pre-SSS X-ray emission can be achieved.

The SSS phase is the brightest phase of X-ray emission
(at constant bolometric luminosity), making it
the best time to observe a nova. We found two clear detections of novae in this
phase. With these two new SSS spectra, the number of Galactic novae clearly
observed in this state is seven \citep[][and reference therein]{orio01,drake03,v4743}.
 For V574\,Pup we were
even able to identify the transition from an early emission line spectrum
to a bright SSS spectrum within two months. V574\,Pup entered its SSS phase
within 250 days after outburst. This is consistent with the evolution of
V1974 Cyg, V1494\,Aql, and V4743\,Sgr (see Fig.~\ref{oday}). The detection
of a SSS in V723\,Cas suggests that nuclear burning is still going on more
than 11 years after outburst. It is not known how much longer it will remain
a SSS. More detailed discussions on V723\,Cas will be presented by Ness et al.
(in prep).

We detected V4743\,Sgr two years after \xmm\ had already measured a very low level
of X-ray emission (see Sect.~\ref{supext}). 
This shows that the post-turn-off evolution can last a few years.
V1494\,Aql was not detected,
but five years prior to our observation it had been
found to have marginal emission and no SSS spectrum.

 In order to compare the evolution of the novae
detected in our sample we plot their distance-scaled X-ray
brightnesses as a function of individual time after outburst
in Fig.~\ref{oday}. In the bottom panel we show the X-ray light curves
(in the same time units) of five more Galactic novae, but observed by
different missions. The brightness is given as unabsorbed luminosity
as extracted from direct measurements of fluxes
\citep{greiner_lmc95,Orio2001,balm98,v4743,petz05}.
As a general rule, any X-ray detection of novae earlier than 100 days after
outburst arises from the hard, shock-generated phase that can either be
emission lines or possibly bremsstrahlung emission \citep{Orio2001,mukai01}. 
Likewise, any X-ray detections more than 1000 days after outburst come from
the nebular phase, with the three exceptions of GQ\,Mus, LMC\,1995, and now
V723\,Cas. These general rules also hold for novae in M31
\citep[see figure 3 in ][]{pietsch05}. 

Finally, we can use the parameters derived from the model fits to the SSS
spectra of V574\,Pup and V723\,Cas to check the \swift\ counts that we
predicted from our attenuated Cloudy model of an ``average'' classical
nova ejection in Fig.~\ref{tvsn} (Sect.~\ref{nh}).
From our blackbody fits we obtained effective temperatures of
$\sim 3\times 10^5$\,K (25\,eV) for both novae, while the derived column
densities were 2.7 and $1.7\times 10^{21}$\,cm$^{-2}$ for V574\,Pup and
V723\,Cas, respectively. With these parameters we find from the grid of
Cloudy models (Fig.~\ref{tvsn}) the predicted counts in a 1-ksec
observation at a distance of 1\,kpc of $\sim 10$ for V574\,Pup and $\sim
100$ for V723\,Cas. Scaling these predicted counts by the distances in
Table~\ref{novaparameters} and the exposure times in Table~\ref{obstab}
(t$_{exp}$(ksec)/D(kpc)$^2$) gives 75 counts for V723\,Cas and 7 counts
for the 30 July, 2005 observation of V574\,Pup. These values are
consistent with the detections given the uncertainties in the
distances and the use of a generalized nova model
\citep[{\it e.g.} a luminosity of $10^{38}$\,erg\,s$^{-1}$,
an ejected mass $\sim$ 10$^{-4}$ M$_{\odot}$,][]{schwarz07}.
For ``normal'' classical novae, calculations
like those done for Fig.~\ref{tvsn} can be used as a tool to
plan X-ray observations by
providing first-order estimates of expected X-ray brightness
levels during the SSS phase.

\acknowledgments

We acknowledge the use of public data from the \swift\ data archive.
We acknowledge with thanks the variable star observations from the AAVSO
International Database contributed by observers worldwide and used in this
research.
J.-U. N. gratefully acknowledges support provided by NASA through Chandra
Postdoctoral Fellowship grant PF5-60039 awarded by the Chandra X-ray Center,
which is operated by the Smithsonian Astrophysical Observatory for NASA under
contract NAS8-03060.
S. Starrfield received partial support from NSF and NASA grants to ASU.
JPO acknowledges support from PPARC

{\it Facilities:} \facility{Swift (XRT)}, \facility{XMM}, \facility{CXO}

\bibliographystyle{apj}
\bibliography{cn,jn,astron,rsoph}

\begin{appendix}

\section{Appendix}

\subsection{Targets}
\label{atargets}

 In this section we summarize some background information of our targets in
addition to that given in Table~\ref{novaparameters}, collected mostly from
unreferenced literature and private communications.

\subsubsection{V574\,Pup}

\object[V574 Pup]{V574\,Pup} was independently discovered by Tago and Sakurai
\citep{2004IAUC.8443....1N}. The evolution of its early light curve
and spectral energy distribution is provided in \cite{2005IBVS.5638....1S}.
The FWHM of H$\alpha$ in the early spectrum was
650\,km\,s$^{-1}$ but with P-Cygni absorption extending to 860\,km\,s$^{-1}$.
Based on the intrinsic colors of the nova both at maximum and at t$_2$, the
same authors derived an extremely low reddening value of E(B--V)$\sim0.05$.
However, this value seems too low due to the low galactic latitude
($b\sim 2\arcdeg$) and the distance ($\sim 3.5$\,kpc) of this nova.
Optical and near-IR spectra obtained one year after outburst showed that
V574\,Pup had entered a coronal phase with lines of [\ion{S}{8}], [\ion{S}{9}],
[\ion{Si}{6}], [\ion{Si}{7}], and [\ion{Ca}{8}] \citep{2005IAUC.8643....2R}.
Analysis of the ratio of the \ion{O}{1} lines in this data set and those
obtained later implies a higher reddening of E(B--V)$= 1.27$
(Rudy, private communication).

\subsubsection{V382\,Nor}

\object[V382 Nor]{V382\,Nor} was discovered prior to visual maximum on
13.3 Mar, 2005 \citep{liller05a}. Examining the AAVSO light curve,
visual maximum probably occurred on 19 March, 2005 at V$\sim9$.
The B--V color obtained on 20 March
was $0.8\,\pm\,0.11$ \citep{liller05a}, which gives an E(B--V) of
$0.57\,\pm\,0.17$ based on the
intrinsic color at maximum \citep[B--V $= +0.23\,\pm\,0.06$; ][]{vandenbergh87}.
Another determination of the reddening uses the intrinsic (B--V) color
at t$_{2}$, $-0.02\,\pm\,0.04$ \citep{vandenbergh87}.
The (B--V) color at t$_{2}$ was $\sim1.1$, which implies E(B--V)$\sim1.1$. This high
a reddening estimate is supported by the saturated \ion{Na}{1} interstellar
lines in the spectra taken at the same time \citep{2005IAUC.8497....2E}.
The P-Cygni profiles seen in the early spectra imply an average expansion velocity
of 1100\,km\,s$^{-1}$.

\subsubsection{V1663\,Aql}

\object[V1663 Aql]{V1663\,Aql} was discovered on 9.2 June, 2005
during routine All Sky Automated Survey patrols
\citep{2005IAUC.8540....1P}. Spectroscopy
obtained one day after maximum showed an extremely red continuum implying
significant reddening toward the source \citep{2005IAUC.8544....1D}.
The emission lines had narrow P-Cygni profiles with expansion velocities of
order 700\,km\,s$^{-1}$.
By 14 November V1663\,Aql had entered its nebular phase with [\ion{O}{3}]
in the optical and the coronal lines of [\ion{S}{8}], [\ion{S}{9}],
[\ion{Si}{6}] and [\ion{Si}{7}] present in the near-IR \citep{2005IAUC.8640....2P}.
The expansion velocity, as measured from the FWHM of the emission lines,
was 2000\,km\,s$^{-1}$ \citep{2005IAUC.8640....2P}. The ratio of the
\ion{O}{1} lines also confirmed
the large extinction, with a value as large as E(B--V)$ = 2$
\citep{2005IAUC.8640....2P}. \citet{lane06} reported an angular
expansion of 0.2\,mas/day at 2.2\,$\mu$m with the Palomar Testbed
Interferometer. This corresponds to a distance of $5.5\,\pm\,1$\,kpc
assuming expansion velocities between $700 - 1000$\,km\,s$^{-1}$.
If the expansion velocities are larger, as measured by \cite{2005IAUC.8640....2P},
then the distance must also be larger than that derived by \citet{lane06}.

\subsubsection{V5116\,Sgr}

On 4 July, 2005 Liller discovered \object[V5116 Sgr]{V5116\,Sgr} already on
the decline. His first observations implied that visual maximum was brighter
than 8 mag \citep{liller05b}. The P-Cygni line profile of H$\alpha$ suggested
an expansion velocity of 1300\,km\,s$^{-1}$ from spectra taken a day after
discovery \citep{2005IAUC.8559....2G}. They also reported a B--V color of
$+0.47\,\pm\,0.02$ on 5 July. Using the intrinsic (B--V) color at maximum for
V5116 Sgr implies an E(B--V)$= 0.24\,\pm\,0.08$.
 
\subsubsection{V1188\,Sco}

V1188 Sco was discovered in late July 2005 \citep{pojmanski05a}.
The visual maximum was measured by \cite{Cooper05}. The
only spectral information in the literature was obtained within
days after visual maximum. Optical spectra were consistent with a
typical CO type nova with broad H$\alpha$, FWHM and Full Width at Zero
Intensity (FWZI) of 1730
and $4000$\,km\,s$^{-1}$, and \ion{Fe}{2} P-Cyg emission lines
\citep{Naito05,Walter05}. A $3-14$ micron spectrum taken one day after
visual maximum displayed a featureless continuum consistent with the
Rayleigh-Jeans tail of the Planck function \citep{Sitko05}. The
\ion{Na}{1} D line was observed with three absorption components
and an equivalent width of 0.5\,nm \citep{Walter05} implying a
large extinction.

\subsubsection{V1047\,Cen}

Little is known about \object[V1047 Cen]{V1047\,Cen} due to its extreme
southern declination ($>-62\arcdeg$). It was discovered by
\cite{liller05c} on 1.031 September, 2005. Visual maximum probably
occurred three days later at V$ = 8.83$ but there is very little
reported data on the early light curve, so its subsequent
behavior, including its t$_2$ decay time, is unknown. The 
spectrum at maximum resembled that of V5114\,Sgr and V5116\,Sgr
with an H$\alpha$ FWHM of $840\,\pm\,50$\,km\,s$^{-1}$ \citep{liller05c}.

\subsubsection{V476\,Sct}

\object[V476 Sct]{V476\,Sct} was discovered after visual maximum on 30.5 Sept,
2005 by A. Takao \citep{2005IAUC.8607....1S}. Haseda (2005) reported an
independent discovery on the same day with a visual magnitude of 10.9.
Early optical
spectra showed many emission lines, notably of \ion{Fe}{2} \citep{Munari06b}.
In the spectra the prominent lines showed double-peak profiles with
velocity separations of $\sim 700$\,km\,s$^{-1}$ and FWHM of order
1000\,km\,s$^{-1}$. Optical and near-IR spectra obtained six weeks after
maximum showed similar emission features but with a strong red continuum
implying the presence of a dust shell in the ejecta. The reddening inferred
from the ratio of the \ion{O}{1} lines at that time was E(B--V)$\sim2$
\citep{2005IAUC.8638....1P}. This value is similar to that derived by
\citet{Munari06b} using the color evolution and strength of the diffuse
6614\,\AA\ interstellar band.

\subsubsection{V477\,Sct}

The second nova discovered in the constellation Scutum in 2005,
\object[V477 Sct]{V477\,Sct}, was detected independently by G. Pojmanski
and K. Haseda on images obtained on the 11th and 13th of October, 2005,
respectively, and the maximum visual magnitude was recorded on 13.07 October
at $V=10.44$\,mag \citep{pojmanski05b}. Early $JHK$ spectroscopy of V477 Sct
showed broad \ion{H}{1} emission lines with FWZI of 6000\,km\,s$^{-1}$ and
no evidence of dust emission \citep{pojmanski05b}. Optical spectra confirmed
broad emission lines with a FWHM value of H$\alpha=2900$\,km\,s$^{-1}$
\citep{Fujii05}. The optical spectra of \cite{Munari06a} showed that unlike
V476\,Sct which was rich in \ion{Fe}{2} lines, V477\,Sct had He and N lines.
Although sparse, the light curve revealed a rapid decline with an estimated
t$_2$ of only 3 days \citep{Munari06a}.
This rapid decline was consistent with the large expansion velocities.
Optical and near-IR spectra taken on 15.094 November, 2005 showed a continuum
increasing toward the red, possibly due to thermal dust emission
\citep{mazuk05}. Numerous emission lines were observed which had
developed double peak profiles with a FWHM of 2700\,km\,s$^{-1}$. The
features present included the coronal lines [\ion{S}{7}] and [\ion{S}{9}] but
not [\ion{Ca}{7}] or [\ion{Si}{6}]. The ratio of the \ion{O}{1} lines
implied a substantial reddening of E(B--V)$= 1.2$ \citep{mazuk05}.
A similar value of $\ge 1.3$ was determined by \citet{Munari06a} based on the
early color evolution, extinction maps along the line of sight, and the large
equivalent width of the diffuse interstellar band at 6614\,\AA.

\subsubsection{LMC\,2005}

Liller discovered \object[Nova LMC 2005]{LMC\,2005} on 26.16 November
2005, but it was also present on images that he had taken four days earlier.
Visual maximum occurred 27 November at $V=12.6$\,mag \citep{liller05d}.
The light curve evolution was extremely slow and in early 2006 exhibited
photometric evidence of dust formation (F. Walter, private communication).

\subsubsection{V723\,Cas}

\object[V723 Cas]{V723\,Cas} is a slow nova. It was discovered on
24 August, 1995 by Yamamoto \citep{1995IAUC.6213....1H} at $V = 9.2$\,mag.
Prediscovery observations showed that it had slowly brightened over the
previous 20 days. The subsequent light curve was unusual. V723 Cas reached a
maximum magnitude of $V\sim8.6$\,mag one hundred days after discovery. At that
point V723\,Cas exhibited a flare to 7.1\,mag on 17 December, 1995. During the
flare the $U-B$ colors became significantly bluer while the B--V color
remained constant \citep{1996A&A...315..166M}. The light curve also had
additional secondary flares during the following 400\,days. V723\,Cas has been 
extensively observed from the ultraviolet \citep{1996IAUC.6295....1G} to radio
\citep{2005MNRAS.362..469H} wavelengths. Recent optical and NIR spectra
showed that V723\,Cas reached a "coronal" phase with emission from high
ionization stages such as [\ion{Fe}{10}] (6373\,\AA) similar to spectra
obtained of GQ Mus in the late 1980s \citep{1989ApJ...341..968K}.
The spectral similarity suggests ongoing nuclear burning
on the WD \citep{1989ApJ...341..968K}. 
As a result, we requested \swift\ observations in order to search for the
X-ray emission that would be related to ongoing nuclear burning.

\subsubsection{V1494\,Aql}

\object[V1494 Aql]{V1494\,Aql} was discovered on 1.8 December, 1999 at
$V = 6.0$\,mag by \cite{pereira99}. Within a few days of discovery this nova
reached a peak magnitude exceeding 4. It subsequently declined by two
magnitudes in $6.6\,\pm\,0.5$ days thus classifying V1494\,Aql as a fast
nova \citep{kissth00}. Three \chandra\,ACIS-I observations were carried 
spectrum with no obvious soft component. However, a bright SSS spectrum
was present in the August spectrum \citep{starf00}.
One month later, two grating observations were obtained
on 28 September and 1 October, 2000, which showed remarkable variability
\citep{drake03}. The spectrum resembles that of the SSS Cal\,83
\citep{paerels01}, except that it appeared somewhat hotter and the
''emission features'' were at different wavelengths. A final observation
with \chandra\ LETGS one year later (ObsID \dataset[ADS/Sa.CXO#obs/2681]{2681})
showed no emission in the dispersed spectrum but a reasonable detection in
the zeroth order on the HRC-S detector. We extracted 56 counts on a background
of 10, corresponding to a count rate of 2\,cts/ksec and a 14-$\sigma$ detection
likelihood. Unfortunately, the HRC detector does
not provide the energy resolution to construct a spectrum.

\subsubsection{V4743\,Sgr}

\object[V4743 Sgr]{V4743\,Sgr} was discovered by \cite{kats}.
A distance of 6.3\,kpc is assumed based on infrared
observations \citep{lyke}. \chandra\ grating observations were carried
out on 19 March, 2003 \citep{v4743}. The light curve showed large
amplitude oscillations ($\sim 20$\% of the count rate) with a period of
1324\,sec (almost 40\% power) with the 2nd and 3rd harmonic overtones present.
V4743\,Sgr was observed by \chandra\ on five more dates and
details of the spectral evolution will be presented by
Ness et al. (in prep).

\subsection{Extraction of source counts}
\label{extraction}

 In this section we give a detailed account of our methods to determine the
count rates and detection probabilities.
 We determined the positions of the sources on the chip from the individual sky
coordinates and used circular extraction regions (radius 10\,pixels$=23.6\arcsec$,
which encircles $\sim 80.5$\,percent of the total PSF). For the extraction of the
background we defined annular extraction regions around the source position with
an inner radius of 10\,pixels and an outer radius of 100\,pixels. We carefully
checked to make sure that no sources were present in the background extraction
regions. We scaled the number of extracted background counts by the ratio of
extraction regions (99.0) to determine the number of expected background counts
per detect cell. We assume Poissonian noise and thus apply a maximum
likelihood estimation in order to determine the net source count rate.
We define two model parameters $S$ and $B$ that represent the number of expected
counts of source and background for the source extraction region
and calculate the number of expected counts in the source- and background
extraction regions $N_s$ and $N_b$:
\begin{equation}
\label{model}
N_s=\alpha S+B\hspace{1cm} N_b=(1-\alpha)S+\beta B
\end{equation}
with $\alpha=0.80$ the fraction of the PSF included in the detect cell and
$\beta=99$ the ratio of background- to source extraction areas.
Under the assumption of Poissonian statistics we can calculate the probability
of finding the measured numbers of counts $n_s$ and $n_b$ in the respective
extraction regions when $S$ and $B$ are given:
\begin{equation}
P=\frac{N_s^{n_s}}{n_s!}e^{-N_s}\frac{N_b^{n_b}}{n_b!}e^{-N_b}
\end{equation}
and thus the likelihood can be calculated
\begin{eqnarray}
\label{like}
{\cal L}= -2\ln P=-2n_s\ln(\alpha S+B) -(\alpha S+B)\\ \nonumber
 +n_b\ln((1-\alpha)S+\beta B) - ((1-\alpha)S+\beta B) +const\,.
\end{eqnarray}
We seek solutions for $S$ and $B$ where ${\cal L}$ reaches a minimum, which
holds for $N_s=n_s$ and $N_b=n_b$, and thus
\begin{equation}
\label{solution}
S=\frac{n_b-\beta n_s}{1-\alpha-\alpha\beta},\ \ B=\frac{n_s-\alpha (n_s+n_b)}{1-\alpha-\alpha\beta}
\end{equation}
leading to a minimum value ${\cal L}_{\rm min}$ of
\begin{equation}
{\cal L}_{\rm min}= -2[n_s(\ln n_s-1)+n_b(\ln n_b-1)]\,.
\end{equation}
With the definition of ${\cal L}$ the range between ${\cal L}_{\min}$ and
${\cal L}_{\min}+1$ is equivalent to the 68.3-percent uncertainty range of the
critical parameter $S$. We thus obtained the formal 1-$\sigma$ errors by varying
$S$ until the value of ${\cal L}$ had increased by 1.0 from the minimum while
leaving $B$ fixed at the value obtained from Eq.~\ref{solution}.

 The value of ${\cal L}_{\rm min}$ can also be used to calculate the detection
probability (in percent) based on the likelihood ratio test, which is a statistical
test of the goodness-of-fit between two models. A relatively more complex model is
compared to a simpler model (null hypothesis) to see if it fits a particular dataset
significantly better. We define the likelihood for the null hypothesis ${\cal L}_0$ as
${\cal L}(S=0)$, and
\begin{equation}
\label{deltal}
\Delta {\cal L}={\cal L}(S)- {\cal L}_0
\end{equation}
quantifies the improvement in ${\cal L}$ after including $S>0$. The parameter $B$
is an "uninteresting parameter" as defined by \cite{avni76}, and we thus convert
$\Delta {\cal L}$ to the probability for {\em one}
degree of freedom using the IDL\footnote{Interactive Data Language, ITT Corporation}
function {\tt chisqr\_pdf}.

In cases where no counts were found in the detect cell ($n_s=0$), or when the number of
counts in the detect cell is smaller than that expected from the background ($n_s<n_b$),
we calculated the 95-percent
upper limits from the number of expected background counts and those actually
measured in the detect cell. We again used the likelihood ratio test and solved
Eq.~\ref{deltal} for ${\cal L}(S)$ according to Eq.~\ref{like} yielding the value of
$S$ that returns $\Delta {\cal L}=4$. We also calculated upper limits in cases where
the derived 1-$\sigma$ uncertainties were larger than the measured count rate.

In addition to the formal detection likelihood
we consider whether the photons inside the source extraction region show
some concentration towards the center, and whether their energy distribution
is different from that of the instrumental background.
In order to assess the concentration towards the center, we give the percentage of
counts that we found in the central quarter of the detect cell in
the last column of Table~\ref{obstab}.
From the PSF we expect 60.5\,percent of source photons to be within a
circle of radius 5\,pixels, and any number much below this fraction is likely not a
source.
We also studied the spectral energy distribution of the background compared to that
of the source as an auxiliary criterion. We computed the median of recorded energies
of the source counts and compared with the same value from the background. These
values are marked by vertical lines (gray dotted for background and black dashed for
source in Figs.~\ref{firstphot} to \ref{lastphot}).
In order to assess the probability that the background spectrum is softer than
the source spectrum, we generated 1000 spectra with $n_s$ counts drawn as random
sub-samples out of the pool of extracted background counts $n_b$ and calculated the
median value for each generated spectrum. While a situation in which
50\,percent of all random cases returned lower median energies does not imply
a non-detection, any strong deviation from 50\,percent is supportive of an
independent source spectrum.

\end{appendix}

\end{document}